\documentclass[12pt,twoside]{article}
\usepackage{amsmath}
\usepackage{amssymb}
\usepackage[T1]{fontenc}
\usepackage{ae,aecompl}
\usepackage{wasysym}
\usepackage{psfrag}
\usepackage{epsf}
\usepackage{graphicx,epsf,amsmath}  
\usepackage{epsf,graphicx}
\newcommand{\alglist}{
\begin{list}{Step 1}
{\setlength{\leftmargin}{0.6 in}\setlength{\labelwidth}{1.0 in}}
}
\title{{\bf Numerical computation of travelling breathers in Klein-Gordon chains}}
\author{Yannick SIRE\thanks{Corresponding author},\ 
Guillaume JAMES \\
~\\
Math\'ematiques pour l'Industrie et la Physique, UMR CNRS 5640\\
~\\
D\'epartement GMM,
Institut National des Sciences Appliqu\'ees,\\
135 avenue de Rangueil, 31077 Toulouse Cedex 4, France.\\
{\small e-mail : 
sire@insa-toulouse.fr, james@insa-toulouse.fr
}
}
\date{October 2004}

\begin{document}
\maketitle
\begin{abstract}
We numerically study the existence of travelling breathers in Klein-Gordon chains, 
which consist of one-dimensional networks of
nonlinear oscillators in an anharmonic on-site potential,
linearly coupled to their nearest neighbors. 
Travelling breathers are spatially localized solutions 
having the property of being exactly translated by $p$ sites along the chain
after a fixed propagation time $T$ (these solutions  
generalize the concept of solitary waves for which $p=1$).
In the case of even on-site potentials, the existence of
small amplitude travelling breathers superposed on a small oscillatory tail
has been proved recently (G. James and Y. Sire, to appear in {\sl Comm. Math. Phys.}, 2004),
the tail being exponentially small with respect to the central oscillation size.
In this paper we compute these solutions numerically and continue them into
the large amplitude regime for different types of even potentials. We find that
Klein-Gordon chains can support highly localized travelling breather solutions
superposed on an oscillatory tail. We provide examples where the tail
can be made very small and is difficult to detect at the scale of central oscillations.
In addition we numerically observe the existence of these solutions in the case of
non even potentials.
\end{abstract}


\section{\label{intro}Introduction}
We consider a chain
of nonlinear oscillators described by the following system 
(discrete Klein-Gordon equation)
\begin{equation} \label{eq:KG}
\frac{d^2 x_n}{d\tau^2}+V'(x_n)=\gamma(x_{n+1}+x_{n-1}-2x_n), 
\,\,n\in \mathbb{Z}. 
\end{equation}
System (\ref{eq:KG}) describes an infinite 
chain of particles linearly coupled to their 
nearest neighbors, in the local potential $V$. 
We denote by
$x_n$ the displacement of the $n$th particle from an equilibrium position, 
$V(x)$ a smooth anharmonic on-site potential satisfying
$V^\prime (0)=0$, $V^{\prime\prime} (0)=1$,
and $\gamma >0 $ the coupling constant.
This class of systems has been used in particular as a prototype for
analyzing the effects of nonlinearity and spatial discreteness
on localization of vibrational energy and energy transport 
(see \cite{review} for a review).
Equation (\ref{eq:KG}) yields in particular
the Peyrard-Bishop model for DNA denaturation \cite{pey,dauxpey}
if one chooses $V$ as a Morse potential.

This paper deals with solutions of (\ref{eq:KG}) satisfying 
\begin{equation} \label{def}
x_n(\tau)=x_{n-p}(\tau-T),
\end{equation}
for a fixed integer $p\geq 1$ ($p$ being the smallest possible) and $T
\in \mathbb{R}$. The case when $p=1$ in (\ref{def}) corresponds to
travelling waves with velocity $1/T $. Solutions satisfying (\ref{def})
for $p\neq 1$ consist of pulsating travelling waves, which are exactly
translated by $p$ sites after a fixed propagation time $T $ and are
allowed to oscillate as they propagate on the lattice. Solutions of type 
(\ref{def}) having the additional property of spatial localization 
($x_{n}(\tau )\rightarrow 0$ as $n\rightarrow \pm \infty $) are known as 
\textit{exact} travelling breathers (with velocity $p/T $) for $p\geq 2$ and
solitary waves for $p=1$.

\textit{Approximate} travelling breather solutions 
(i.e. spatially localized solutions satisfying (\ref{def}) only approximately)
have been extensively studied. 
These solutions have
been numerically observed in various one-dimensional nonlinear lattices such
as Klein-Gordon chains \cite{dauxois,ting,chen,aubryC,cuevas2},
Fermi-Pasta-Ulam lattices \cite{takeno,bickham,sandusky,flachW} 
and the discrete nonlinear Schr\"odinger (DNLS) equation \cite{flach,eilbeck}
(see \cite{review} for more references). 
These solutions have been studied also in more sophisticated nonlinear 
models describing
energy transport in alpha-helix \cite{hyman,scott} and DNA \cite{hennig}.
One way of generating approximate
travelling breathers consists of ``kicking'' static breathers consisting of
spatially localized and time periodic oscillations (see the basic paper \cite{aubryMK}
for more details on these solutions). Static breathers are put into motion
by perturbation in the direction of a pinning mode \cite{chen,aubryC}. The
possible existence of an energy barrier (``Peierls-Nabarro barrier'')
that the breather has to overcome in
order to become mobile has drawn a lot of attention, see e.g. 
\cite{dauxois,aubryC,flachW,kastner} and the review paper 
\cite{sepulchre}.
An analytical method recently developed by MacKay and Sepulchre \cite{mackay}
allows to approximate the nonuniform breather motion along the chain
(the breather center approximately moves under the action of an effective potential).
An application to the Fermi-Pasta-Ulam lattice can be found in \cite{kastner}.

In the small amplitude regime, spatially localized solutions of
(\ref{eq:KG})-(\ref{def})
have been studied analytically by Iooss and Kirchg\"assner \cite{ioossK} for $p=1$
(case of travelling waves). 
Travelling wave solutions of (\ref{eq:KG}) have the form 
$x_n(\tau )=x_0 (\tau -n T )$ and are determined by the scalar advance-delay
differential equation 
\begin{equation}  \label{scal}
\frac{d^2 x_0}{d\tau^2}+V^\prime (x_0)= \gamma \, (x_0 (\tau +T) -2x_0 +x_0
(\tau -T)).
\end{equation}
Extension to larger values of $p$ 
has been performed by James and Sire \cite{sirejames,jamessire,sireprep}
(case of pulsating travelling waves).
For fixed $p\geq 1$, problem (\ref{eq:KG})-(\ref{def}) reduces to the $p$-dimensional
system of advance-delay differential equations
\begin{eqnarray}\label{systemp} 
\frac{d^2}{d\tau^2}\left[
\begin{array}{c}
    x_1\\
\vdots \\
x_n \\
\vdots \\
    x_p
  \end{array}\right]+
\left[
\begin{array}{c}
    V'(x_1)\\
\vdots \\
 V'(x_n) \\
\vdots \\
    V'(x_p)
  \end{array}\right]=\gamma
\left[
\begin{array}[c]{c}
    x_2(\tau)-2x_1(\tau)+x_p(\tau+T)\\
\vdots \\
x_{n+1}(\tau) -2x_n (\tau) +x_{n-1}(\tau)\\
\vdots \\
    x_1(\tau-T)-2x_p(\tau)+ x_{p-1}(\tau)
  \end{array}\right] .
\end{eqnarray}
System (\ref{systemp}) is analyzed 
in the above references
using a center manifold reduction in the infinite
dimensional case, as described e.g. in reference \cite{ioossV}. 
It is rewritten as a reversible evolution
problem in a suitable functional space, and considered for parameter values 
$(T ,\gamma )$ near critical curves
where the imaginary part of the spectrum of the linearized operator consists
of a pair of double eigenvalues and pairs of simple ones
(the number of pairs is finite and equals the number of resonant phonons). 
Close to these curves, the pair of double eigenvalues splits 
into two pairs of hyperbolic
eigenvalues with opposite nonzero real parts, which opens the possibility of
finding solutions homoclinic to $0$.
Near these parameter values, the center manifold theorem reduces the problem
locally to a reversible finite dimensional system of differential equations.
The reduced system is put in a normal form which is integrable up to higher
order terms. In some regions of the parameter space, the 
{\it truncated} normal
form admits reversible orbits homoclinic to $0$, which bifurcate from the
trivial state and correspond to approximate solutions of (\ref{systemp}).

Exact small amplitude
solutions of (\ref{eq:KG})-(\ref{def}) close to these approximate solutions
(in a sense that we shall specify, allowing the existence of a small
oscillatory tail) have been obtained for $p=1$ \cite{ioossK}, and for
$p=2$ in the case of even potentials $V$ \cite{jamessire}.
In both cases,
the simplest homoclinic bifurcation yields a
$6$-dimensional reversible reduced system.
As it is shown by Lombardi for different classes of reversible
systems having a slow hyperbolic part and fast oscillatory modes
\cite{lombardi} (see also \cite{champneys} for related numerics),
reversible solutions of the truncated normal form
homoclinic to $0$ should not \textit{generically}
persist when higher order terms are taken into account in the normal form.
The existence of corresponding travelling waves 
($p=1$) or travelling breathers ($p=2$)
decaying exactly to $0$
should be a codimension-$1$ phenomenon, the codimension depending on the
number of pairs of simple purely imaginary eigenvalues
in the considered parameter regime (there is one
pair of purely imaginary eigenvalues, in addition to hyperbolic ones).
However, to confirm the nonexistence of reversible orbits homoclinic to $0$
(close to a small amplitude homoclinic orbit of the truncated normal form) 
\emph{for a given choice of $V$, $\gamma$, $T$}, one has to check the
nonvanishing of a certain Melnikov function being extremely difficult to
compute in practice \cite{lombardi}.
Due to this codimension-$1$ character, in a given system (\ref{systemp}) 
(with $p=1$, or $p=2$ and $V$ even), $\gamma$ and $V$ being fixed, 
exact travelling wave or travelling breather
solutions decaying to $0$ at infinity might exist in the small
amplitude regime, but for isolated values of the velocity $p/T$
(see \cite{flachZK} for examples of nonlinear
lattices supporting explicit travelling breather solutions with
particular velocities). 

Instead of orbits homoclinic to $0$, the full normal form admits 
orbits homoclinic 
to small periodic ones \cite{lombardi} (originating from the pair of
purely imaginary eigenvalues). These solutions correspond to exact solitary
wave ($p=1$) or travelling breather ($p=2$)
solutions of (\ref{eq:KG}) superposed on a small periodic oscillatory
tail, which can be made exponentially small with respect to the central
oscillation size (the minimal tail size should be \textit{generically}
nonzero for a given value of $(T ,\gamma )$). This phenomenon is
in accordance with numerical observations in the case of
travelling waves \cite{aubryC}, and should explain numerical convergence
problems arising in the computation of fully localized
solitary waves or travelling breathers
in various nonlinear lattices \cite{sze,abl}.

For asymmetric potentials $V$ and $p\geq 2$, 
the simplest homoclinic bifurcation yields a
higher-dimensional ($2p+4$-dimensional) reduced system, with supplementary
pairs of simple imaginary eigenvalues of the linearized operator (the
imaginary part of the spectrum consists of a pair of double eigenvalues and
$p$ pairs of simple ones). 
By analogy with results of Lombardi \cite{lombardi}, it has been
conjectured \cite{jamessire,sireprep} that 
the above mentioned solutions of the truncated normal form
homoclinic to $0$ do not 
\textit{generically} persist when higher order terms are taken into account in the
normal form. Persistence might be true if parameters ($T ,\gamma $, 
coefficients in the Taylor expansion of $V$) are chosen on a
discrete collection of codimension-$p$ submanifolds of the parameter space. For
general parameter values, instead of orbits homoclinic to $0$ one can expect
the existence of reversible orbits homoclinic to exponentially small $p-$%
dimensional tori, originating from the $p$ additional pairs of simple purely
imaginary eigenvalues. These solutions should constitute the principal part
of exact travelling breather solutions of (\ref{eq:KG}) superposed on a
small quasi-periodic oscillatory tail. However, in order to obtain exact
solutions one has to prove the persistence of the corresponding homoclinic
orbits as higher order terms are taken into account in the normal form. This
step is non-trivial and would require to generalize results of Lombardi 
\cite{lombardi} available for a single pair of simple imaginary eigenvalues 
(another promising approach is developed in the recent work \cite{il}). 

One might think that
this nonpersistence picture contrasts with the case of the
integrable Ablowitz-Ladik lattice \cite{ladik}, which supports
an explicit family of exact travelling breather solutions
in which the breather velocity can be varied continuously. 
However these solutions are not robust under various
non-Hamiltonian reversible perturbations as shown in \cite{berger}.

In this paper we numerically solve system (\ref{systemp}) in the case 
$p=2$, which constitutes the simplest situation for travelling breather
solutions (the case $p=1$ corresponding to travelling waves). 
This allows us to go beyond the analytical results obtained
in \cite{jamessire}, which were restricted to the small amplitude regime.

Our numerical observations confirm
that system (\ref{eq:KG})-(\ref{def}) (with $p=2$)
admits travelling breather solutions superposed on a small
oscillatory tail.
These solutions are either weakly or strongly localized depending on
the value of $T$ ($\gamma$ being fixed).
These results are obtained both for hard and soft on-site potentials.
In addition we provide examples where the tail's amplitude
can be made extremely small near particular values of the breather velocity,
which makes the tail difficult to detect at the
scale of central oscillations. 
Note that travelling breather solutions have been previously computed 
in the DNLS equation \cite{feddersen,ablowitz}
(in fact gauge symmetry reduces the problem to the computation of travelling waves), 
but the possible existence of a very small oscillatory tail has not been addressed.

An intuitive explaination for the existence of a tail
is that travelling breathers transfer
energy to resonant phonons. The energy transfer can be balanced by 
the superposition of a resonant phonon tail, which allows the
breathers to propagate in an exact way.
This remark might suggest more generally that a way to enhance breather mobility
in nonlinear oscillator chains
could be to excite resonant phonons with
appropriate amplitudes.

Our numerical computations are performed as follows.
We consider system (\ref{systemp}) for $p=2$ and rescale time 
by setting $t=\frac{\tau}{T}$. We introduce 
the variable $(u_1(t),u_2(t))=(x_1(\tau),x_2(\tau+\frac{T}{2}))$, 
which is a solution of the symmetrized system
\begin{eqnarray} \label{eq:system}
\frac{d^2}{dt^2}\left[
\begin{array}[c]{c}
    u_1\\
    u_2
  \end{array}\right]+
T^2\left[
\begin{array}[c]{c}
    V'(u_1)\\
    V'(u_2)
  \end{array}\right]=\gamma T^2
\left[
\begin{array}[c]{c}
    u_2(t-\frac{1}{2})-2u_1(t)+u_2(t+\frac{1}{2})\\
    u_{1}(t+\frac{1}{2})-2u_2(t)+u_1(t-\frac{1}{2})
  \end{array}\right] 
\end{eqnarray} 
and is related with the original variables by the relations
\begin{eqnarray}
\label{xn}
&x_{n}(\tau)=u_1(\frac{\tau}{T}-\frac{n-1}{2})\,\, \mbox{if $n$ is odd, }\\
&x_{n}(\tau)=u_2(\frac{\tau}{T}-\frac{n-1}{2})\,\, \mbox{if $n$ is even.}\nonumber
\end{eqnarray}
We solve system (\ref{eq:system}) by finite difference discretization and
a Newton method (the Powell hybrid method \cite{powell}). 
An alternative would be to use spectral methods, as done in
\cite{eilbeckF,duncan,savin,rothos} for the computation of
travelling waves.

In addition, we derive a simpler problem in the case of 
even potentials $V$. In this case, 
system (\ref{eq:system}) admits symmetric solutions
satisfying $u_1=-u_2$, corresponding to solutions of (\ref{eq:KG}) having
the form 
\begin{equation}
\label{xnscal}
x_n(\tau)=(-1)^{n+1}u_1(\frac{\tau}{T}-\frac{(n-1)}{2}).
\end{equation}
This yields a new scalar advance-delay differential equation 
\begin{equation}\label{scalar}
\frac{d^2u_1}{dt^2}+T^2V'(u_1)=-\gamma T^2 (u_1(t+1/2)+2u_1(t)+u_1(t-1/2)).
\end{equation}
The convergence of the Newton method heavily relies on having a good
initial guess. For this purpose we use 
the explicit principal parts of solutions
provided by center manifold theory in the small amplitude regime \cite{jamessire},
and continue them numerically by varying $T$. 

An alternative numerical method would be to consider equation
(\ref{eq:KG}) and  search for fixed points of the time-$T$ map
$F\, : \, (x_n(0),\dot{x}_n(0))\mapsto (x_{n+p}(T),\dot{x}_{n+p}(T))$,
which correspond to solutions satisfying (\ref{def}).
This method has been used in \cite{aubryC} for $p=1$ (case of travelling waves).
The first step of the method consists in computing a static breather,
which is then perturbed in the direction of a pinning mode in order
to obtain a (slowly moving) approximate travelling breather.
This solution serves in turn as an initial guess for the
Newton method, as one computes fixed points of $F$. 
One limitation of the method is that pinning modes do not always exist
(this is the case e.g. for certain hard quartic on-site potentials),
but the method yields very precise results at least for travelling waves \cite{aubryC}.

The paper is organized as follows. Section \ref{sec1} describes
the numerical scheme and suitable initial guesses for the Newton method. 
Section \ref{sec2} is devoted to the scalar case (\ref{scalar}). 
System (\ref{eq:system}) is numerically solved in section \ref{sec3}.

\section{\label{sec1}Numerical method}
\subsection{Finite difference scheme}
Equations (\ref{eq:system}) and (\ref{scalar}) have 
the following general form 
\begin{equation}\label{general}
\frac{d^2U}{dt^2}+W(U)=\sum_{j=-1}^1 A_jU(t+\frac{j}{2}),
\end{equation}
where $U(t)\in \mathbb{R}^d$ ($d=1,2$)
and $A_{-1},A_0,A_1$ are $d \times d$ matrix. The nonlinear function $W:\mathbb{R}^d \rightarrow \mathbb{R}^d$ is smooth and satisfies $W(0)=0$. 
We consider periodic boundary conditions $U(t+M)=U(t)$.
For large $M$ one can expect to find good approximations
to spatially localized solutions which have ``infinite period''
(in practice we shall fix $M=20$ most of the time).

We shall approximate (\ref{general}) using a 
second order finite difference scheme. We choose the interval of discretization $I=[-\frac{M}{2},\frac{M}{2}]$ and a discretization step $h=\frac{1}{2N}$ where $M,N\in \mathbb{N}^*$. 
We fix an integer period $M$ so that time delays
in (\ref{general}) will be multiple of the step size, but
this feature is not essential (see section \ref{peror}).

We set $U_i=U(hi)$ with $hi \in I$ and $i=-MN+1,...,MN$. The finite difference scheme writes
\begin{equation}\label{num}
\frac{U_{i+1}-2U_i+U_{i-1}}{h^2}+W(U_i)=\sum_{j=-1}^1 A_jU_{i+jN},
\end{equation}
for $i=-MN+1,...,MN$. We impose the following periodic boundary conditions
\begin{equation}\label{BC}
U_{i+2MN}=U_i.
\end{equation}
Equations (\ref{num})-(\ref{BC}) lead to a nonlinear system of $2MN$ equations with $2MN$ unknowns $U_i$. We use a Powell hybrid method \cite{powell} to solve this system 
(we impose a stoping criteria at precision of $10^{-12}$).

\begin{figure}[h]
\psfrag{tau}[0.9]{\vspace{0.05cm}{\small $\ln h$}}
\psfrag{amp}{{\small $\ln E$}}
\begin{center}
\includegraphics[scale=0.21]{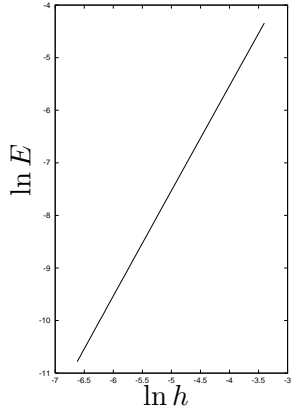}
\end{center}
\caption{\label{erreur} Relative error 
$E=\| \{ x_n (T)-x_{n-2}(0) \}  \|  \, /\, \| \{ x_n (T) \}  \|$
($\| \, \|$ denotes the Euclidean norm)
as a function of step $h$ (in logarithmic scales), 
for a travelling breather solution at $T=7.7$, 
$\gamma\approx 0.9$, with the polynomial potential $V(x)=\frac{1}{2}x^2-\frac{1}{4}x^4$.} 
\end{figure}

Numerical solutions of (\ref{general}) 
obtained by solving (\ref{num})-(\ref{BC})
are checked by an integration of 
system (\ref{eq:KG}) with $(x_n(0),\dot{x}_n(0))$ as initial condition
($x_n$ is deduced from equation (\ref{xn}) or (\ref{xnscal}), and one considers $2M$ 
lattice sites with periodic boundary conditions). 
The numerical integration is performed with
a $4^{th}$ order Runge-Kutta scheme. 
We impose that $x_n$ is shifted by $2$ sites after integration 
over $[0,T]$ with a relative error tolerance of order $10^{-5}$. 
Figure \ref{erreur} gives a typical profile of
the evolution of the relative error as a function of step $h$ 
(note that the error decays as $h^2$). 

The crucial step 
is the choice of a good inital guess for the Powell method. 
This aspect is examined in next section.

\subsection{\label{inguess}Initial guess}
The existence of small amplitude travelling breather solutions of 
(\ref{eq:KG}) (having a small oscillatory tail) has been proved
in \cite{jamessire} in the case of even potentials, using 
a center manifold technique. 
The method also provides leading order approximate expressions for these solutions 
(these approximate expressions are also available in the case of
asymmetric potentials).
This material is described below.

\begin{figure}[h]
\begin{center}
\includegraphics[scale=0.7]{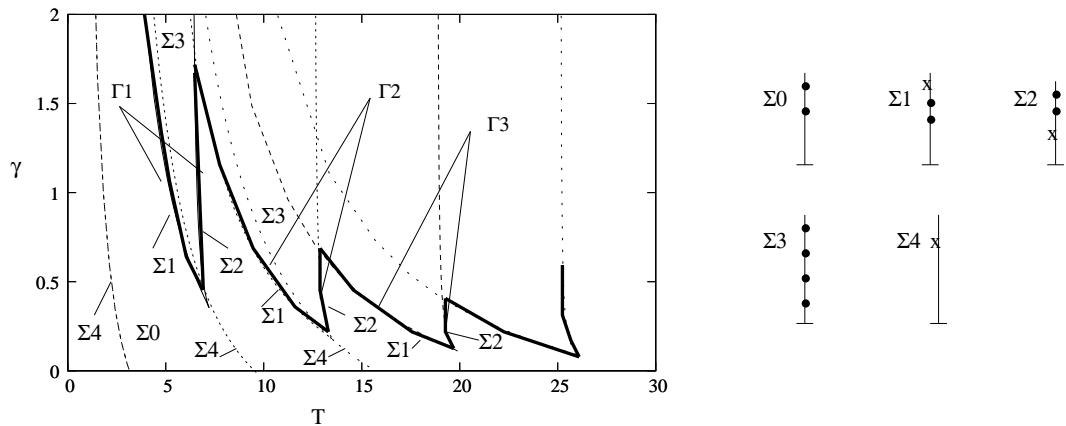}
\end{center}
\caption{\label{bif} 
Sketch of the eigenvalues $iq$ on the upper imaginary axis
(right), in different regions of the parameter space (left).
These eigenvalues correspond to real solutions of (\ref{disp})
(here $q>0$).}
\end{figure}

We consider the linearization of system (\ref{eq:system}) at $(u_1,u_2)=(0,0)$ and search for
solutions of the linearized problem in the form 
$(u_1,u_2)=e^{iq t}(\hat{u}_1,\hat{u}_2)^T$. This leads to the following dispersion relation 
\begin{equation}\label{disp}
(-q^2+T^2(1+2\gamma))^2-4(\gamma T^2)^2\cos^2(q/2)=0.
\end{equation} 
Equation (\ref{disp}) provides the spectrum (consisting in 
eigenvalues $\pm iq$)
of a certain
linearized evolution operator obtained as system (\ref{eq:system})
is rewritten in the form of a reversible differential equation in
a Banach space (see \cite{jamessire} for more details).
The spectrum on the imaginary axis corresponds to solutions
$q\in\mathbb{R}$ of (\ref{disp}) and
is sketched in figure~\ref{bif}. 

Double roots $q\in\mathbb{R}$ of (\ref{disp}) satisfy in addition 
\begin{equation}\label{disp2}
2q(-q^2+T^2(1+2\gamma))=(\gamma T^2)^2 \sin(q) 
\end{equation}
(solutions of (\ref{disp})-(\ref{disp2}) correspond to
eigenvalues of the linearized evolution operator which are
in general double, and at most triple \cite{jamessire}).
We note that for
$T^2(1+2\gamma)=(2k+1)^2\pi^2$ ($k\in \mathbb{N}$), $q=(2k+1)\pi$ 
is a double root of (\ref{disp}). This situation
occurs on the curves denoted as $\Sigma_4$ in figure \ref{bif}, and
corresponds to the crossing of two simple roots of (\ref{disp}) as
these curves are crossed in the parameter plane.
The other double roots defined by system (\ref{disp})-(\ref{disp2})
are found as parameters lie on
the curve $\Gamma\, :\, (T(q),\gamma(q))$ ($q\in \mathbb{R}^+$) 
defined by
\begin{equation}\label{ioossCurve1}
T^2=q^2-4q \tan(q/4),
\end{equation}
\begin{equation}\label{ioossCurve2}
\gamma=\frac{2q}{T^2\sin(q/2)},
\end{equation}
if $q\in [4k\pi,(2k+1)2\pi]$ (for an integer $k\geq 1$), and
\begin{equation}\label{moiCurve1}
T^2=q^2+\frac{4q}{\tan(q/4)},
\end{equation}
\begin{equation}\label{moiCurve2}
\gamma=-\frac{2q}{T^2\sin(q/2)},
\end{equation}
if $q\in [(2k-1)2\pi,4k\pi]$ ($k\geq 1$).
The range of $q$ is determined by the condition $T^2>0$. 
One can observe that $\Gamma$ lies in the parameter region $\gamma T^2 >4$. 

We denote by $\Gamma_k$ the restriction of $\Gamma$ to the interval $q\in [2k\pi ,2(k+1)\pi ]$ (these curves form ``tongues'' as shown in figure \ref{bif}). 
Our numerical computations will start near $\Gamma_1$ where
bifurcation of small amplitude travelling breathers take place. 
Travelling breathers bifurcate more generally
near $\Gamma_{2k+1}$ \cite{jamessire}, 
and bifurcating solitary waves are found
near $\Gamma_{2k}$ \cite{ioossK}.

Double roots on $\Gamma$ correspond to the appearance (or vanishing)
of a pair of real roots of (\ref{disp}) as the curve $\Gamma$ is crossed.
The envelope $\Delta$ of $\Gamma$ is the bold line 
depicted in figure \ref{bif}.
Below it (and outside $\Sigma_4$), real solutions of (\ref{disp})
consist in two pairs of simple roots $\pm q_1$, $\pm q_2$.
An additional pair of real (double) roots $\pm q_0$ appears as
one reaches $\Delta$ from below
(they correspond to pairs of hyperbolic eigenvalues of the
linearized evolution operator colliding on the imaginary axis).
In the sequel, we shall exclude points of $\Delta$
which are close to points where 
$sq_0+rq_1+r'q_2=0$ for $s,r,r' \in \mathbb{Z}$ and $0<|s|+|r|+|r'| \leq 4$
(such values correspond to strong resonances),
and denote this new set as $\Delta_0$.

Now we fix $(T_0,\gamma_0 )\in \Gamma_m\cap \Delta_0$. 
For $(\gamma ,T )\approx (\gamma_0,T_0)$, it has been proved
in \cite{jamessire} that small amplitude solutions of
(\ref{eq:system}) have the form
\begin{equation}
\label{manif}
\left(
\begin{array}[c]{c}
    u_1(t)\\
    u_2(t) 
  \end{array}\right)=
A(t)\left(
  \begin{array}[c]{c}
    (-1)^m\\
    1
  \end{array}\right)
+
C(t)\left(
  \begin{array}[c]{c}
    -1\\
    1
  \end{array}\right)
+
D(t)\left(
  \begin{array}[c]{c}
    1\\
    1
  \end{array}\right)
+\mbox{c.c.}
+
\Psi (u_c (t),\gamma ,T) ,
\end{equation}
where $u_c =(A,B,C,D,\bar{A},\bar{B},\bar{C},\bar{D})^T\in \mathbb{C}^8$
and $\Psi \, : \, \mathbb{C}^8\times \mathbb{R}^2 \rightarrow \mathbb{R}^2$
is a smooth function, with $\Psi (u_c ,\gamma ,T)=
O(\| u_c\|^2 +\| u_c\| (|\gamma -\gamma_0 |+|T-T_0 |)) $ 
as $(u_c ,\gamma ,T)\rightarrow (0,\gamma_0, T_0)$
(i.e. $\Psi$ contain higher order terms).
The complex coordinates $(A,B,C,D)$ satisfy a reversible differential equation
\begin{eqnarray} \label{normalform}
\frac{dA}{dt}=iq_0A+B+iA\mathcal{P}(v_1,v_2,v_3,v_4)\nonumber
+O((|A|+|B|+|C|+|D|)^{4}),\\\nonumber
\frac{dB}{dt}=iq_0B+iB\mathcal{P}(v_1,v_2,v_3,v_4)+A\mathcal{S}(v_1,v_2,v_3,v_4)\\\nonumber
+O((|A|+|B|+|C|+|D|)^{4}),\\
\frac{dC}{dt}=iq_1C+iC\mathcal{Q}(v_1,v_2,v_3,v_4)
+O((|A|+|B|+|C|+|D|)^{4}),\\\nonumber
\frac{dD}{dt}=iq_2D+iD\mathcal{T}(v_1,v_2,v_3,v_4)
+O((|A|+|B|+|C|+|D|)^{4}),\\\nonumber
\end{eqnarray}
where
\begin{equation*}
v_1=A\bar{A},
v_2=C\bar{C},
v_3=D\bar{D},
v_4=i(A\bar{B}-\bar{A}B)
\end{equation*}
and $\mathcal{P},\mathcal{S},\mathcal{Q},\mathcal{T}$ are affine functions 
with smoothly parameters dependent real coefficients, for $(T,\gamma)$ in the neighborhood of $\Delta_0$. 
We have
\begin{eqnarray}\label{poly}\nonumber
\mathcal{P}(v_1,v_2,v_3,v_4)=p_1(\gamma,T)+p_2v_1+p_3v_2+p_4v_3+p_5v_4,\\
\mathcal{S}(v_1,v_2,v_3,v_4)=s_1(\gamma,T)+s_2v_1+s_3v_2+s_4v_3+s_5v_4,\\
\nonumber
\mathcal{Q}(v_1,v_2,v_3,v_4)=\tilde{q}_1(\gamma,T)+\tilde{q}_2v_1
+\tilde{q}_3v_2+\tilde{q}_4v_3+\tilde{q}_5v_4,\\
\mathcal{T}(u_1,u_2,u_3,u_4)=
t_1(\gamma,T)+t_2v_1+t_3v_2+t_4v_3+t_5v_4\nonumber
\end{eqnarray}
where $p_1,s_1,\tilde{q}_1,t_1$ vanish on $\Delta_0$.
The method for computing the coefficients of (\ref{poly}) is
given e.g. in \cite{jamessire}, section 5.2
(see \cite{ioossA} for a description of the method in a general
setting).
Defining the coefficients in the expansion of $V$ as
\begin{equation}
\label{defv}
V(x)=\frac{1}{2}x^2+\frac{\alpha}{3}x^3+\frac{\beta}{4}x^4+O(|x|^5),
\ \ \ x\rightarrow 0,
\end{equation} 
one has in particular for $(T,\gamma )=(T_0,\gamma_0 )\in \Delta_0$
\begin{equation}\label{s2}
(2-\frac{q_0}{\tan(q_0/2)})s_2=T_0^2(-6\beta+8\alpha^2-\frac{4\alpha^2T_0^2}{2\gamma_0T_0^2\cos(q_0)-T_0^2(1+2\gamma_0)+4q_0^2}).
\end{equation}
Moreover, as $(T,\gamma )\rightarrow (T_0,\gamma_0 )\in \Delta_0$ one has 
$$
s_1=s_{11}(\gamma-\gamma_0)+s_{12}(T-T_0)+\mbox{h.o.t},
$$
with
\begin{equation}
\label{coefs1}
s_{11}=-4T_0^2\frac{\epsilon-\cos(q_0/2)}{2\epsilon-\frac{\gamma_0T_0^2}{2}\cos(q_0/2)},\, \, 
s_{12}=4T_0\frac{-(1+2\gamma_0)\epsilon+2\gamma_0\cos(q_0/2)}{2\epsilon-\frac{\gamma_0T_0^2}{2}\cos(q_0/2)}.
\end{equation}
We set $\epsilon=-1$ in (\ref{coefs1}) if 
$q_0\in [(2k-1)2\pi,4k\pi ]$, $k\geq 1$
(i.e. $ (T_0,\gamma_0 )\in \Gamma_{2k-1}$)
and $\epsilon=1$ if $q_0\in [4k\pi ,(2k+1)2\pi]$ 
($ (T_0,\gamma_0 )\in \Gamma_{2k}$).

Equation (\ref{manif}) expresses the fact that small solutions 
of (\ref{eq:system})
lie on a $8$-dimensional center manifold $\mathcal{M}_{\gamma ,T}$, 
the dimension of which is equal
to the number of marginal modes of equation (\ref{eq:system})
linearized at $(u_1,u_2)=0$ for $(\gamma ,T)=(\gamma_0 ,T_0)$.
One can parametrize the center manifold in such a way that
the reduced system
(i.e. the $8$-dimensional differential equation satisfied by the
coordinates $u_c$ of solutions on $\mathcal{M}_{\gamma ,T}$)
is as simple as possible (this form is called ``normal form'').
The normal form of order $3$ of the reduced system is given
in (\ref{normalform}).
The truncated normal form (obtained by neglecting terms of orders $4$ 
and higher in (\ref{normalform})) 
is integrable \cite{ioossP}, \cite{jamessire}.

For $(T,\gamma )\approx(T_0 ,\gamma_0 )\in \Gamma_{m}  \bigcap \Delta_0$ and
if $(T, \gamma )$ lies below $\Delta$ (this corresponds
to having $s_1 >0 $ in (\ref{poly})), the linearized operator at
the right side of (\ref{normalform}) admits two pairs of symmetric
hyperbolic eigenvalues close to $\pm iq_0$, and
two pairs of purely imaginary eigenvalues close to $\pm iq_1 ,\pm iq_2$.
If $s_2(\gamma_0,T_0)<0$, the {\it truncated} normal form
admits reversible homoclinic solutions to $0$ which bifurcate from
the trivial state as $(T,\gamma )\rightarrow (T_0 ,\gamma_0 )$.
They correspond near $\Gamma_{2k+1}$
to {\it approximate} travelling breather
solutions of (\ref{eq:KG}), given by system (\ref{eq:system})
and having the form
\[\left(
\begin{array}[c]{c}
    u_1(t)\\
    u_2(t) 
  \end{array}\right)\approx
A(t)\left(
  \begin{array}[c]{c}
    -1\\
    1
  \end{array}\right)
+\mbox{c.c.}
\]
(this expression comes from the principal parts of (\ref{manif}),(\ref{normalform})), where 
\begin{equation}\label{homsol2}
A(t)=r_0(t)e^{i(q_0t+\psi(t))},\
\end{equation}
and 
\begin{equation}
\label{coefansatz}
r_0(t)=(\frac{2s_1}{-s_2})^{1/2}
(\cosh(t
s_1^{1/2}
))^{-1},\ \ \
\psi(t)=
p_1t+2\frac{p_2}{s_2}
s_1^{1/2}
\tanh(t
s_1^{1/2}
).
\end{equation}

The ansatz (\ref{homsol2}) provides a good inital guess 
for computing small amplitude travelling breather solutions with
the Powell method. The solutions
initially computed are then followed by continuation as a 
function of $T$ towards higher amplitudes. 

The normal form coefficients $p_1,p_2$ in (\ref{coefansatz})
can be computed as explained in \cite{ioossA}.
However, the term $\psi (t)$ in (\ref{homsol2}) 
is negligible as $(T,\gamma )\approx(T_0 ,\gamma_0 )$
since $p_1, s_1 \approx 0$ and $t\in [-\frac{M}{2},\frac{M}{2}]$.
Consequently one can fix $\psi =0$ in (\ref{homsol2})
to initiate the Powell method.

In addition, it is not necessary to 
take oscillatory modes $C(t),D(t)$ into account  
in the initial ansatz (\ref{homsol2}) because
those are conjectured to be exponentially small with
respect to $|\gamma -\gamma_0|+|T-T_0|$ as
$(\gamma, T)\rightarrow (\gamma_0,T_0)$,
for some exact solutions of (\ref{eq:system})
(this has been proved for even potentials \cite{jamessire}).
Note in addition that a tail appears naturally as one iterates
the Powell method.

In the following section, we present numerical computations 
in a particular symmetric case (for even potentials $V$)
in which travelling breather solutions are given by 
a scalar advance-delay differential equation. 
\section{\label{sec2}Case of even potentials $V$}

We consider the case when the potential $V$ in (\ref{eq:KG}) is even. 
In this case, there exist solutions of (\ref{eq:KG}) satisfying 
\begin{equation}
\label{symx}
x_{n+1}(\tau)=-x_n(\tau-\frac{T}{2}), 
\end{equation}
given by the
scalar advance-delay differential equation (\ref{scalar}).
Section \ref{exact} reviews analytical results \cite{jamessire} concerning
small amplitude travelling breather solutions of the form (\ref{symx}). 
These solutions
are numerically computed 
in section \ref{numcomp}
and continued into the large amplitude regime.
In section \ref{peror} we numerically identify
these solutions as modulations of certain time-periodic oscillations
(time-periodic pulsating travelling waves).

\subsection{\label{exact}Exact small amplitude travelling breathers}

In this section we sum up analytical results obtained in \cite{jamessire}, which
follow from the center manifold reduction theorem
and the analysis of the reduced equation (\ref{normalform}).

\begin{figure}[h]
\begin{center}
\includegraphics[scale=0.4]{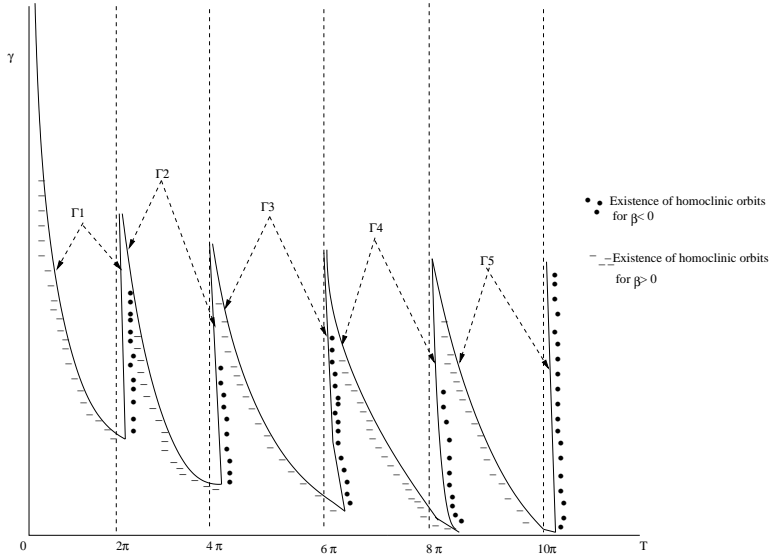}
\end{center}
\caption{\label{a0}Regions in the parameter space where small amplitude
homoclinic orbits to $0$ bifurcate for the truncated normal form
(case of an even potential).}
\end{figure}

The homoclinic orbits of the truncated normal form
described in section \ref{inguess} 
bifurcate near parts of the ``tongues'' $\Gamma_{m}$
where $s_2 <0$. For even potentials ($\alpha=0$ in (\ref{defv})),
homoclinic bifurcations occur near left branches of
$\Gamma_{m}$ if $V$ is hard ($\beta>0$ in (\ref{defv})),  
and near right branches if $V$ is soft ($\beta<0$).
The situation is depicted in figure \ref{a0}
(see \cite{jamessire} for more details).

Now assume $s_2(\gamma_0, T_0)<0$ defined by equation (\ref{s2}) for a fixed 
$(T_0,\gamma_0)\in \Delta_0 \bigcap \Gamma_{2k+1}$ and consider 
$(\gamma,T)\approx (\gamma_0, T_0)$, $(T,\gamma )$ lying below $\Delta$
in the parameter plane.
The full reduced equation (\ref{normalform}) admits an invariant subspace
defined by $D=0$ due to the evenness of $V$ (see \cite{jamessire}).
Considering (\ref{normalform}) on this invariant subspace is equivalent to
searching for solutions of (\ref{eq:KG})-(\ref{def}) satisfying (\ref{symx})
($m=2k+1$ and $u_2 =-u_1$ in (\ref{manif})).

If $(T_0,\gamma_0)$ lies outside some subset of $\Delta_0 \bigcap \Gamma_{2k+1}$ 
having zero Lebesgue measure (corresponding to resonant cases), 
the full reduced equation (\ref{normalform}) admits small amplitude
reversible solutions (with $D=0$) homoclinic to periodic orbits. 
These solutions correspond to {\it exact} travelling breather solutions 
of system (\ref{eq:KG}) superposed at infinity on an oscillatory 
(periodic) tail. Their principal part is given by
\begin{equation}
\label{solappr2}
x_n(\tau)=(-1)^n
[\,
A +
C 
\, ]
\, (\frac{\tau}{T}-\frac{n-1}{2})
+\mbox{c.c.}+\mbox{h.o.t}, 
\end{equation}
where $(A,C)$ is a solution of (\ref{normalform}) homoclinic to
a periodic orbit (see \cite{jamessire} for an approximation of $(A,C)$ at leading order).
For a fixed value of $(\gamma,T)$ (and up to a time shift), these
solutions occur in a one-parameter family parametrized by 
the amplitude of oscillations at infinity.
The lower bound of these amplitudes is
$O(e^{-c/\mu^{1/2}})$, where
$\mu =|T-T_0|+|\gamma -\gamma_0|$, $c>0$.   

The lower bound of the amplitudes
should be {\it generically} nonzero, but may vanish on 
a discrete collection of curves in the parameter plane
$(T,\gamma )$. As a consequence, in a given system (\ref{eq:KG})
(with fixed coupling constant $\gamma$ and symmetric on-site potential $V$),
exact travelling breather solutions decaying to $0$
at infinity (and satisfying (\ref{def}) for $p=2$)
may exist in the small amplitude regime, for isolated values of the
breather velocity $2/T$.

In the following section, we fix $(T_0 ,\gamma_0 )\in \Gamma_1$ 
and consider $\mu=|T-T_0|$ as a small parameter, $\gamma=\gamma_0$ being fixed. 
We numerically solve equation (\ref{scalar}) and follow the above mentioned
travelling breather solutions into the large amplitude regime.  

\subsection{\label{numcomp}Numerical computation of travelling breathers}

\begin{figure}
\psfrag{tau}[0.9]{{\small $t$}}
\psfrag{amp}{{\small$u_1(t)$}}
\psfrag{n}[0.9]{{\small $n$}}
\psfrag{xn}{{\small$x_n(0)$}}
\begin{center}
\includegraphics[scale=0.2]{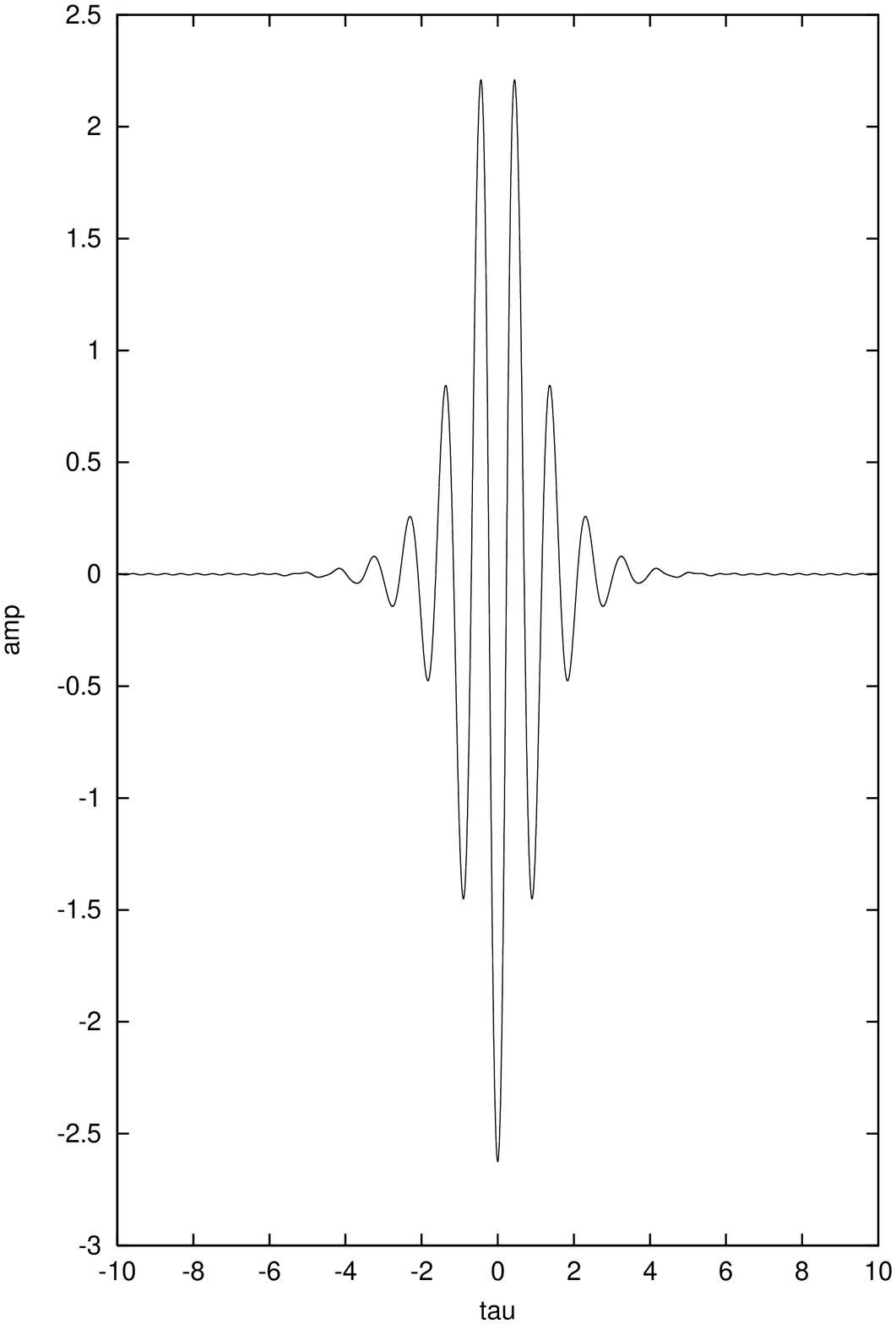}
\includegraphics[scale=0.37]{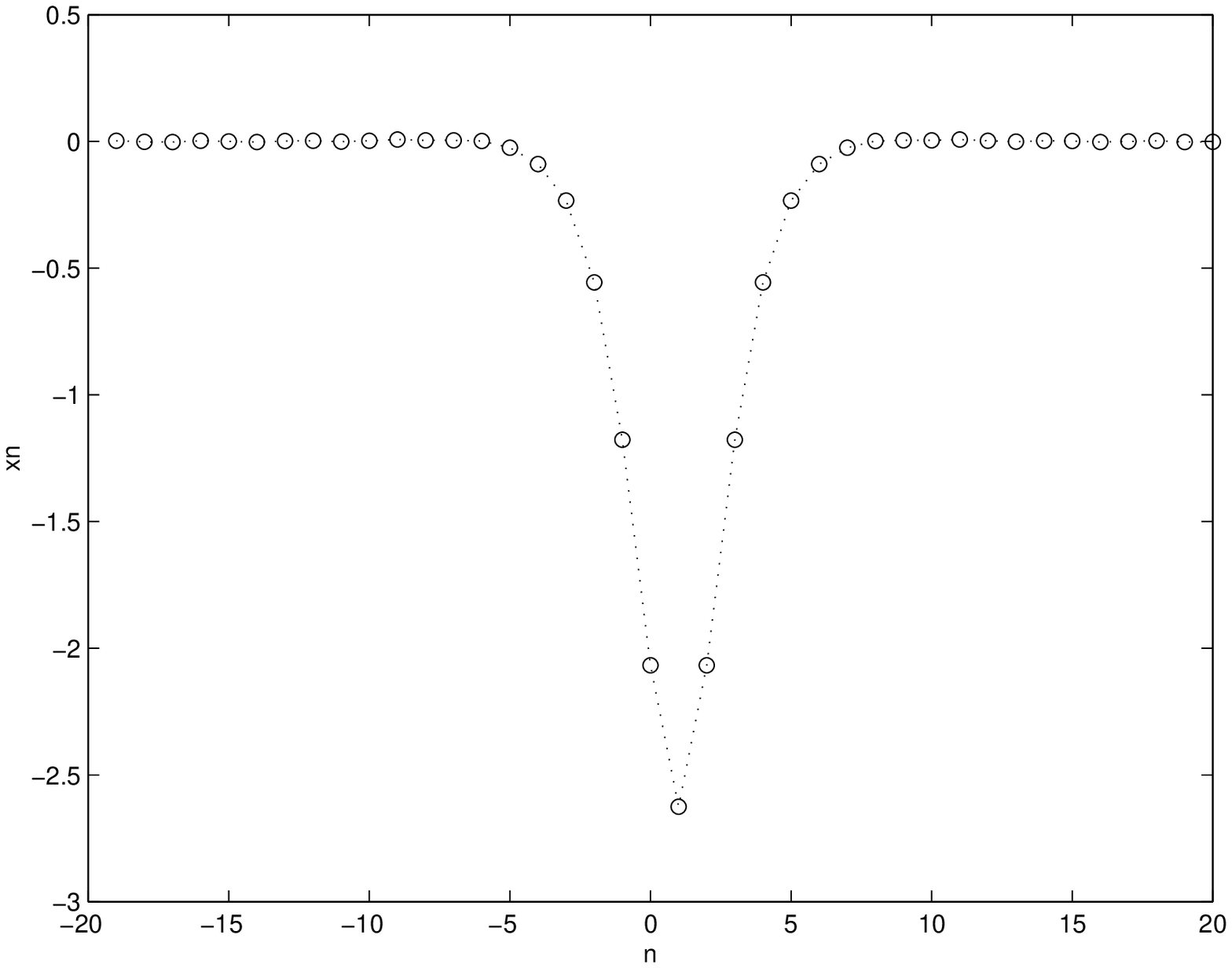}
\end{center}
\caption{\label{SG} Large amplitude solution in the case of the trigonometric potential $V(x)=1-\cos(x)$ with $T=8.1,\gamma=\gamma_0 \approx 0.9$, $T_0 \approx 6.63$ 
($(T_0 ,\gamma_0 ) \in \Gamma_1$). The left figure shows the displacements of mass $n=1$ as a function of $t$ (note that $t$ is a rescaled time). 
The right one shows displacements for all lattice sites at time $\tau =0$.}
\end{figure}

Let us consider the trigonometric potential $V(x)=1-\cos(x)$ 
($\beta=-\frac{1}{6}$ in (\ref{defv})) and fix
$\gamma \approx 0.9$, $T=8.1$. These parameter values lie near the existence
domain of small amplitude travelling breather solutions (see section \ref{exact}).
However, parameters are chosen sufficiently far from
 the curve $\Gamma_1$ on which travelling breathers bifurcate
(the point $(T_0 ,\gamma_0)  \approx (6.63,0.9)$ lies on $\Gamma_1$),
and correspond thereby to highly localized solutions.
Figure \ref{SG} shows a solution $u_1 (t)$ of (\ref{scalar}) (left),
and the displacements of lattice sites at time $\tau =0$ deduced from equation (\ref{xnscal})
(right).

\begin{figure}
\psfrag{tau}[0.9]{{\small $t$}}
\psfrag{amp}{{\small$u_1(t)$}}
\psfrag{n}[0.9]{{\small $n$}}
\psfrag{xn}{{\small$x_n(0)$}}
\begin{center}
\includegraphics[scale=0.2]{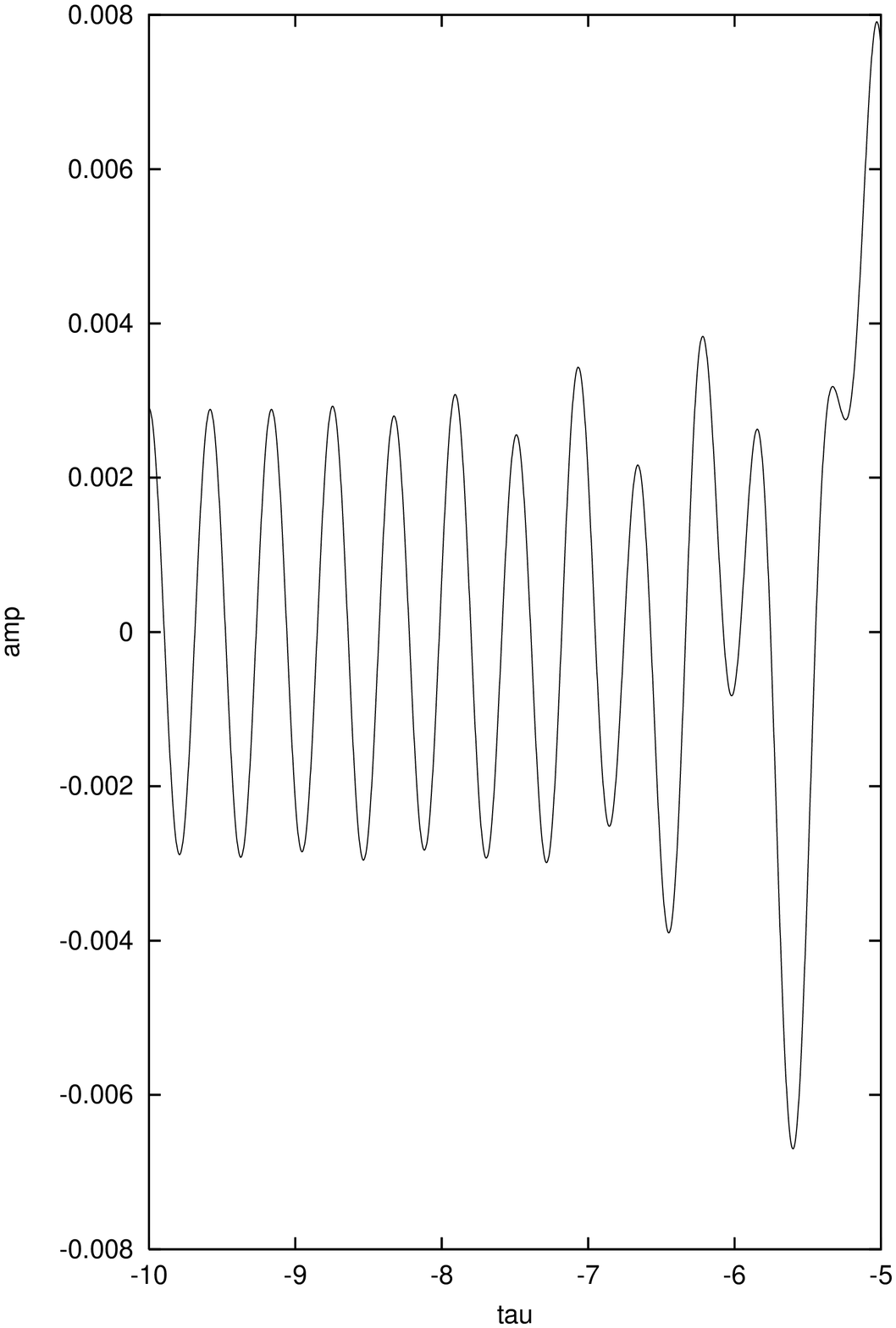}
\includegraphics[scale=0.37]{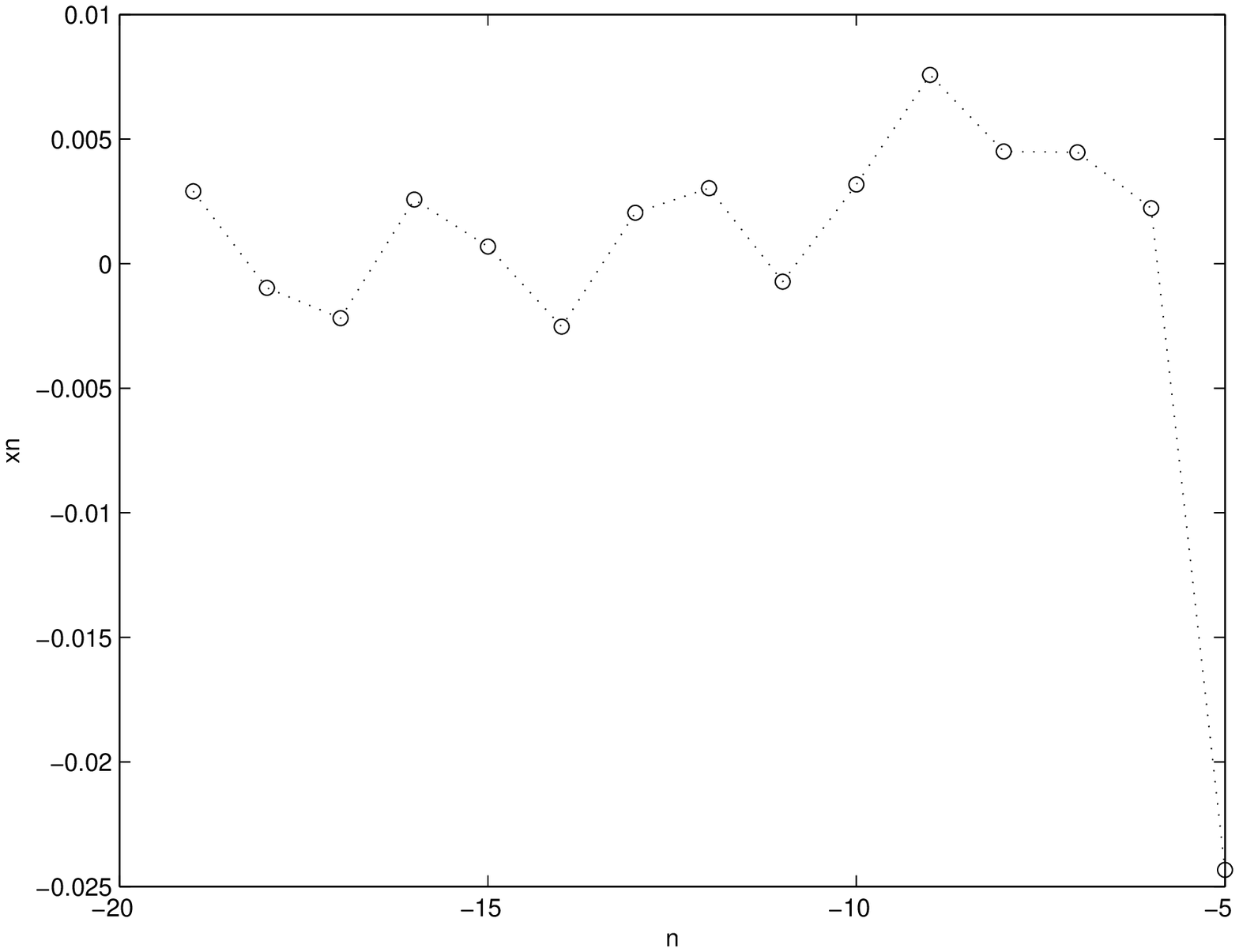}
\end{center}
\caption{\label{zoomtrig} Magnification of the solution tail in figure \ref{SG} (case of a trigonometric potential $V$, $T=8.1$, $\gamma \approx 0.9$).}
\end{figure}

This result shows that the Klein-Gordon lattice with a trigonometric potential
supports highly localized travelling breather solutions.
Although the solution in figure \ref{SG} seems perfectly localized, 
it admits in fact a small oscillatory tail not visible at the figure scale
(this phenomenon is known for small amplitude solutions, see section \ref{intro}).
The tail is clearly visible after magnification
(figure \ref{zoomtrig}).

\begin{figure}
\psfrag{tau}[0.9]{\vspace{0.05cm} {\small $t$}}
\psfrag{amp}{{\small$u_1(t)$}}
\begin{center}
\includegraphics[scale=0.21]{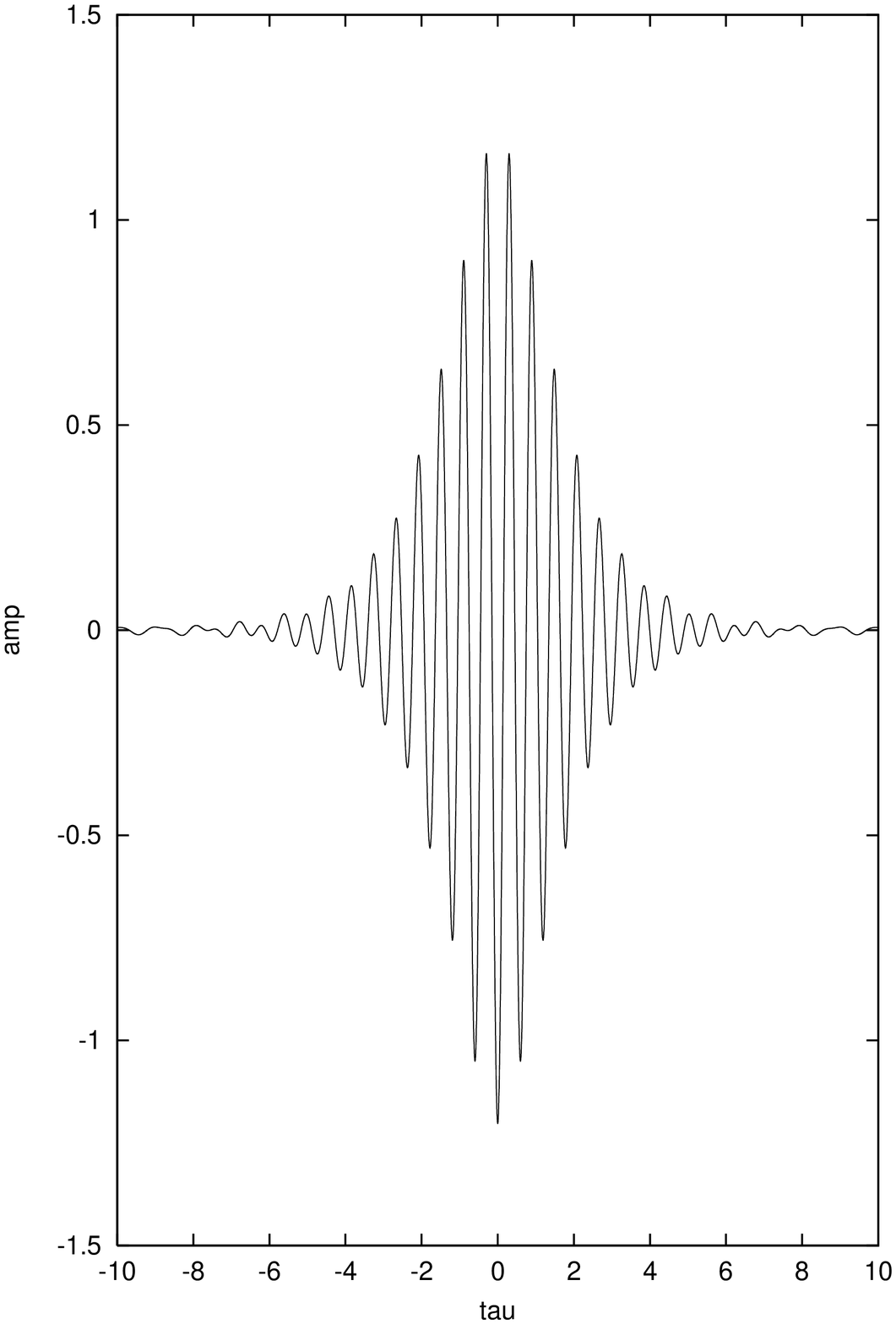}
\includegraphics[scale=0.21]{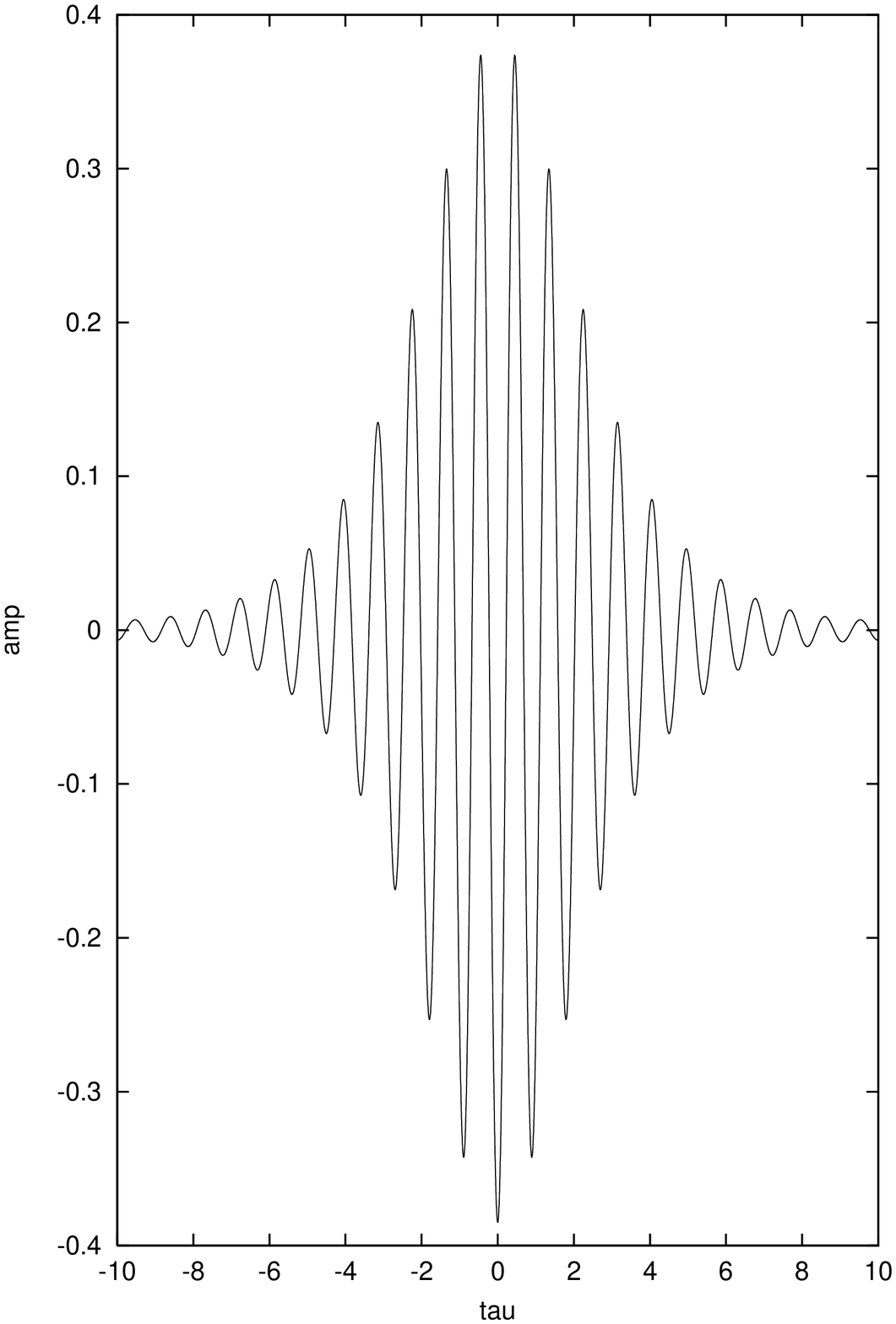}
\end{center}
\begin{center}
\includegraphics[scale=0.21]{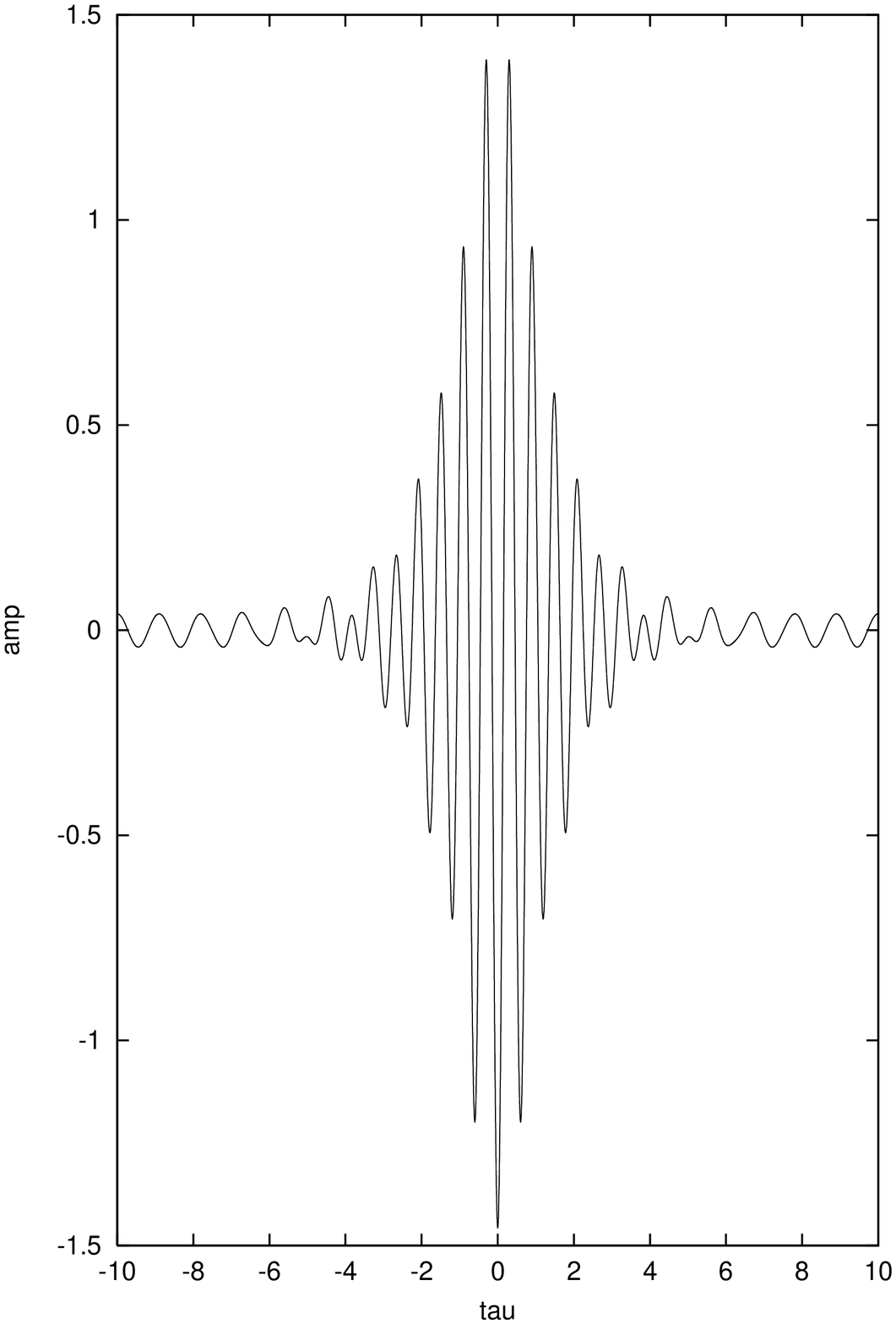}
\includegraphics[scale=0.21]{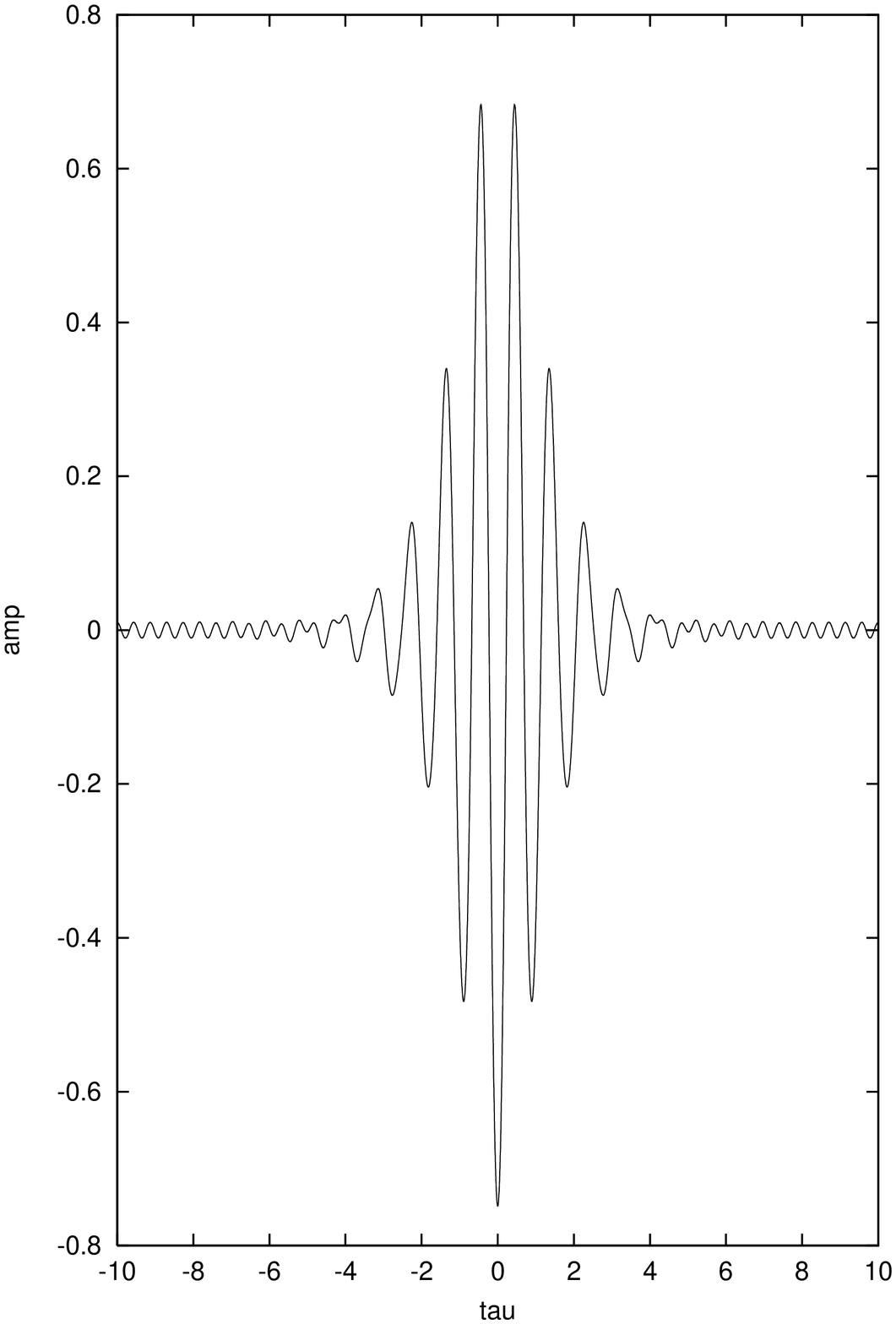}
\end{center}
\begin{center}
\includegraphics[scale=0.21]{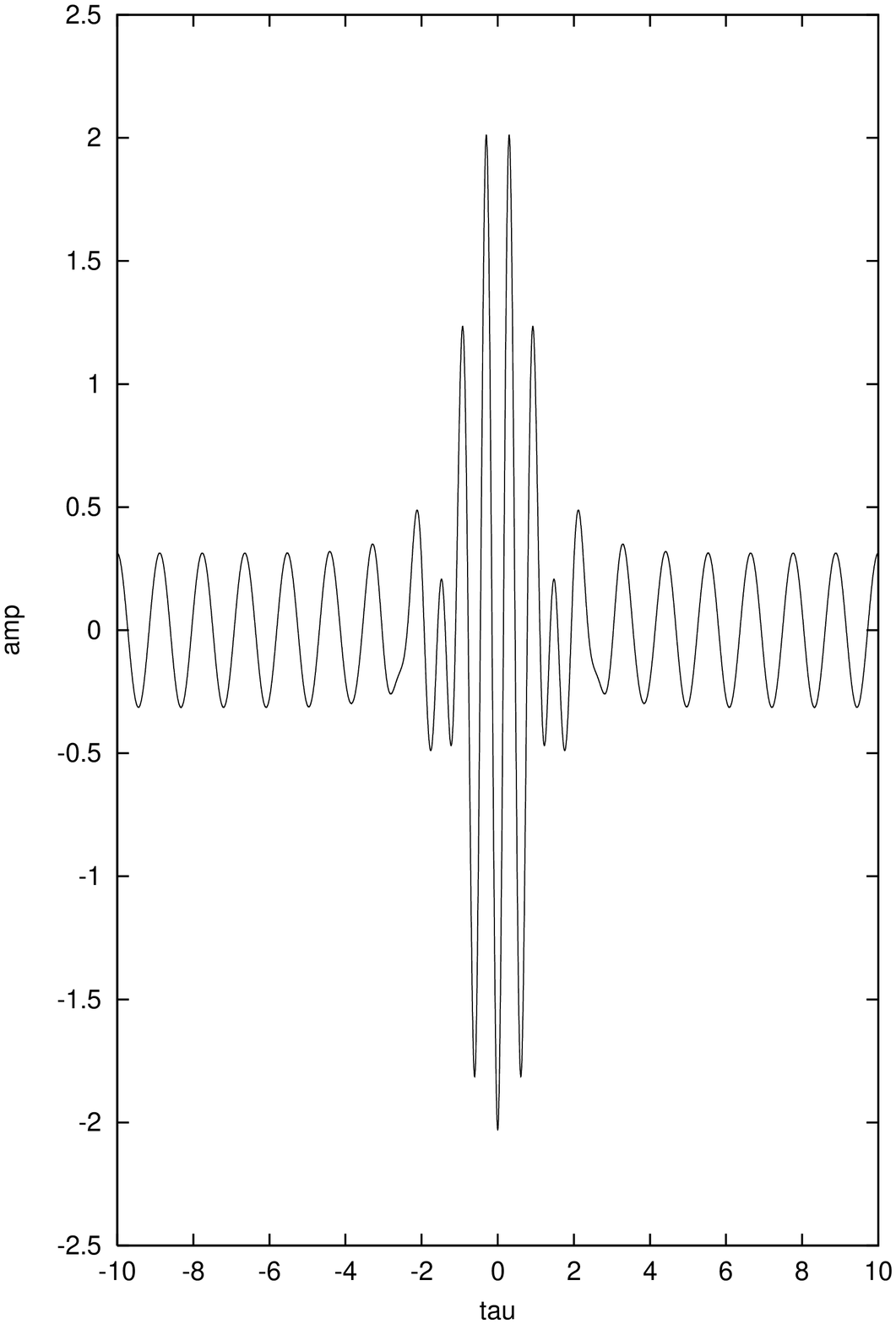}
\includegraphics[scale=0.21]{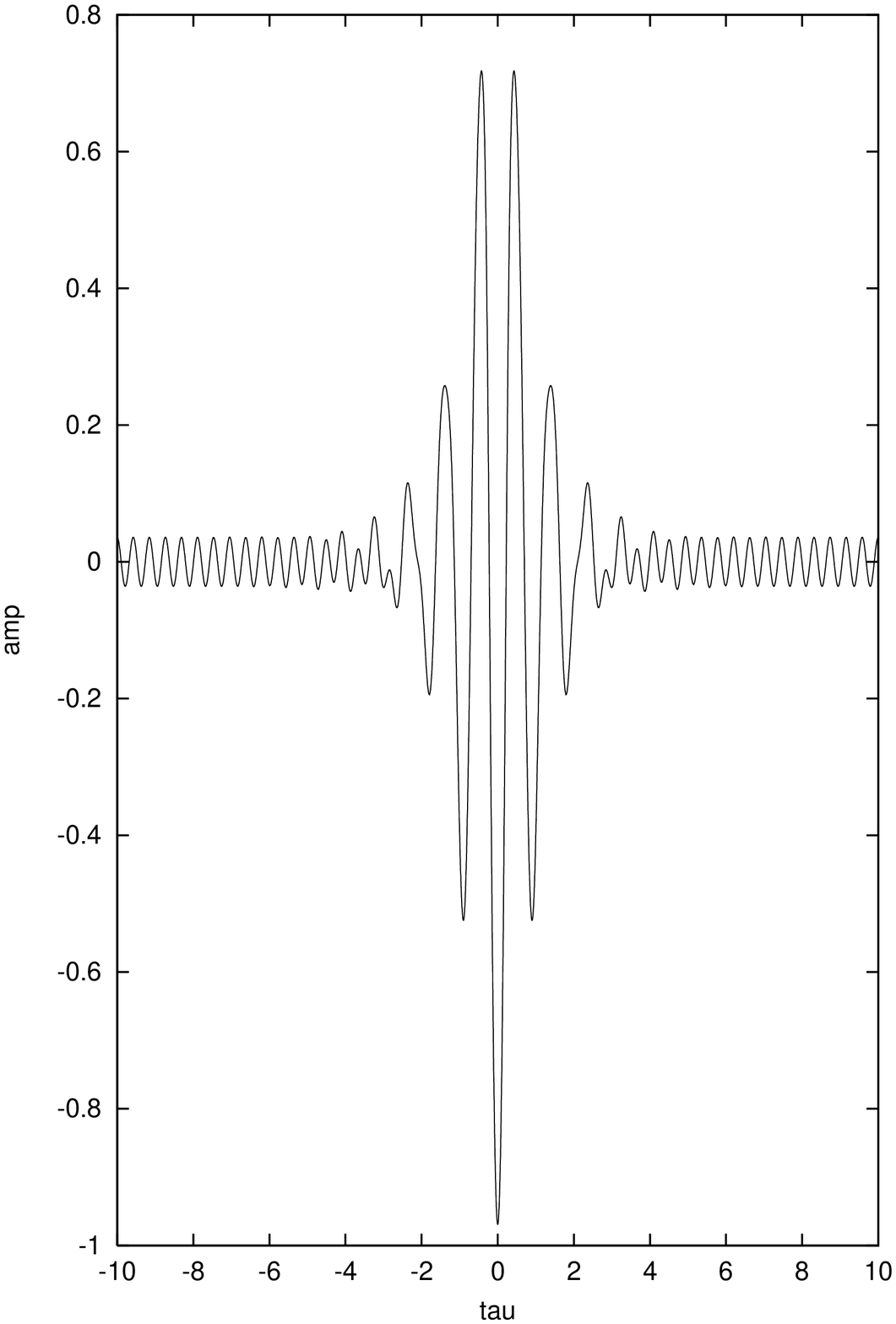}
\end{center}
\caption{\label{solutions} Localized numerical solutions of (\ref{scalar}) for various propagation times $T$ and polynomial potentials $V(x)=\frac{1}{2}x^2+\frac{\beta}{4}x^4$. The left figures correspond to a hard potential 
($\beta=1/6$, $\gamma=\gamma_0 \approx 0.83$, $T_0 \approx 5.59$ and $T=5.53$, $T=5.5$, $T=5.35$ from top to bottom) and the right ones to a soft potential ($\beta=-1$, $\gamma=\gamma_0 \approx 0.9$, $T_0 \approx 6.63$ and $T=6.8$, $T=7.4$, $T=8.1$ from top to bottom). 
The value of $\mu=|T-T_0|$ increases from top to bottom.}
\end{figure}

In the sequel we study in more detail the properties of these solutions as $T$ is
varied, and depending whether one considers a hard or soft on-site potential.
For this purpose we shall fix 
$V(x)=\frac{1}{2}x^2+\frac{\beta}{4}x^4$.
Figure \ref{solutions} presents numerically computed solutions of (\ref{scalar})
for different values of the propagation time $T$. 
Left column corresponds to the hard potential 
$V(x)=\frac{1}{2}x^2+\frac{1}{24}x^4$ ($\beta =1/6$), 
the right one to the soft potential $V(x)=\frac{1}{2}x^2-\frac{1}{4}x^4$
($\beta =-1$). 
We start close to parameter values $(T_0 ,\gamma_0 ) \in \Gamma_1\cap \Delta_0$ 
and fix $\gamma =\gamma_0$.

\begin{figure}
\vspace{1cm}
\psfrag{omega}[0.9]{\small $ q/ (2\pi ) $}
\psfrag{four}{\small$|\hat{u}_1|^2$}
\begin{center}
\includegraphics[scale=0.3]{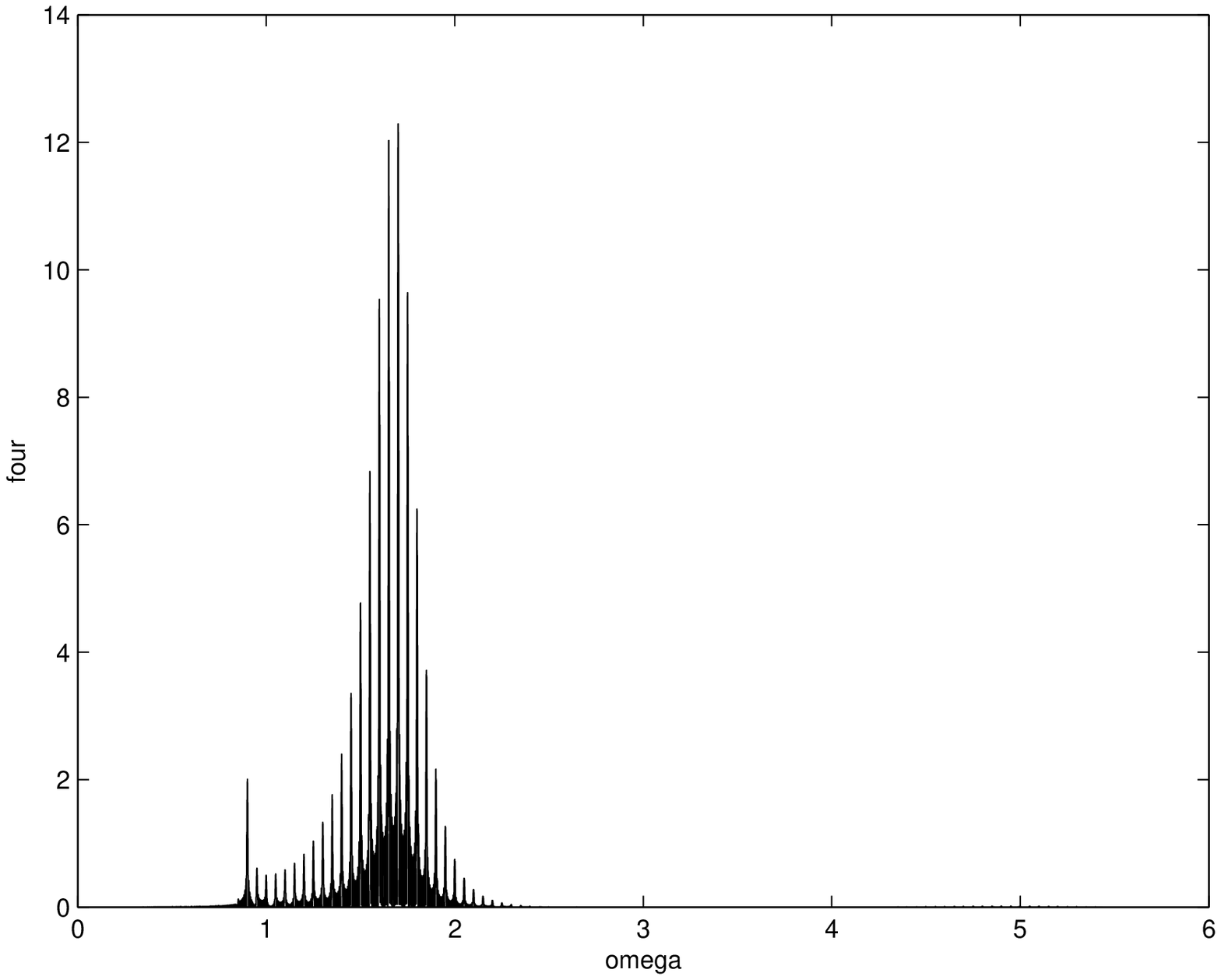}
\includegraphics[scale=0.3]{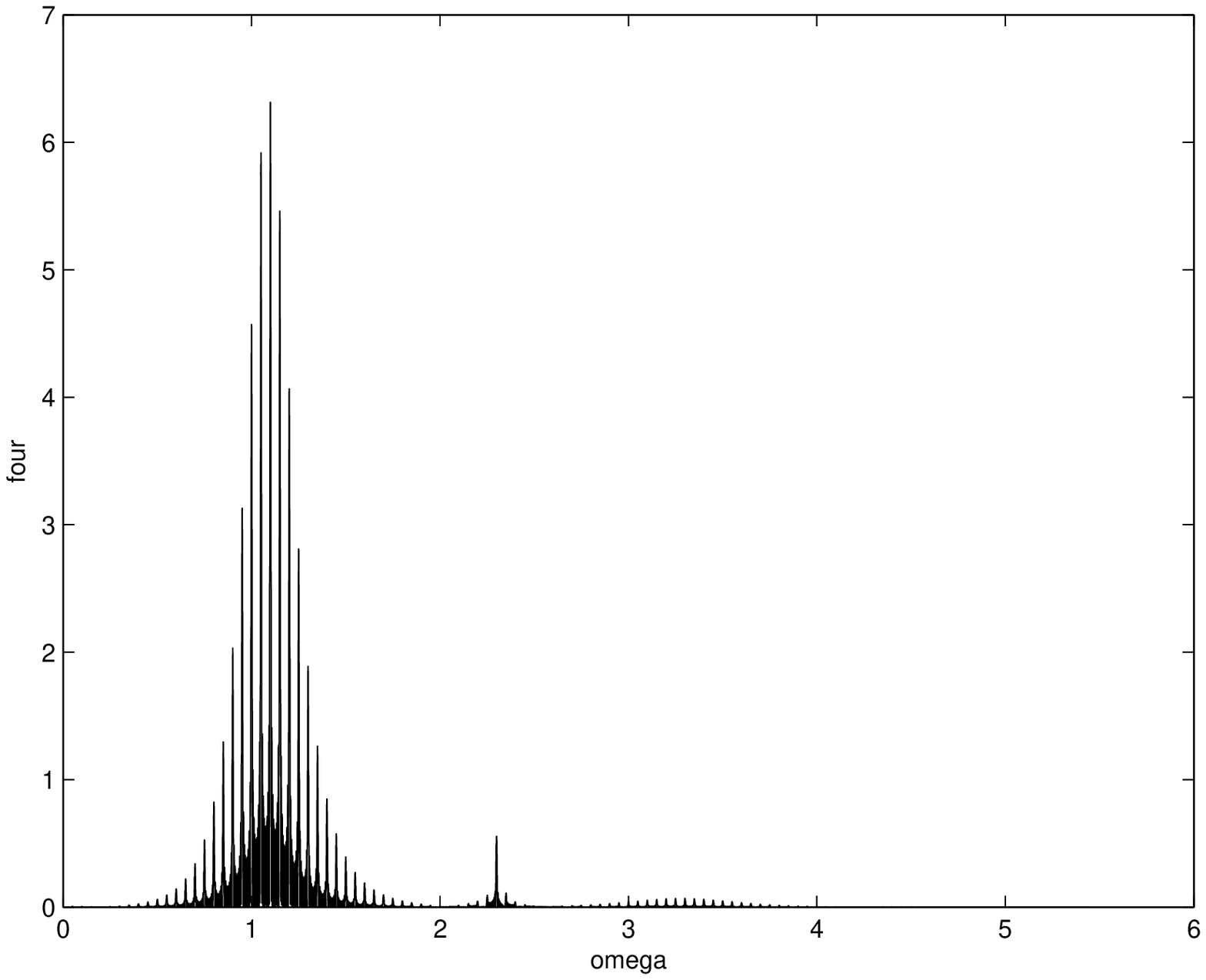}
\end{center} 
\caption{\label{fourier} Fourier spectra of $u_1 (t)$
(profiles of figure \ref{solutions} are extended with periodic
boundary conditions). 
The right figure corresponds to a solution at  
$T=7.4,\gamma=\gamma_0 \approx 0.9$, $T_0 \approx 6.63$ 
for $V(x)=\frac{1}{2}x^2-\frac{1}{4}x^4$
(see figure \ref{solutions}, right). 
The left one corresponds to a solution at  
$T=5.5,\gamma=\gamma_0 \approx 0.83$, 
$T_0 \approx 5.59$ for $V(x)=\frac{1}{2}x^2+\frac{1}{24}x^4$
(see figure \ref{solutions}, left). 
The peaks are relatively close to 
$q_0/2\pi \approx 1.11, q_1/2\pi \approx 2.30$ in the soft potential case 
and $q_0/2\pi \approx 1.7, q_1/2\pi \approx 0.9$ in the hard case.}
\end{figure}

One can check that as $\mu =|T-T_0|$ increases, the central peak of the
solution gets more localized and increases in amplitude. 
In addition, a periodic tail becomes more visible as $\mu$ increases.
One can observe a frequency difference between the central part of solutions and 
their tail, as it is the case in the small amplitude regime
(in equation (\ref{solappr2}), the central part $A(t)$ and the tail $C(t)$
oscillate respectively with frequencies close to the linear frequencies
$q_0$ and $q_1$).
In particular the tail has a lower frequency in the case of hard potentials,
and a larger one for soft potentials.
In figure \ref{fourier} we plot the Fourier spectrum of $u_1 (t)$ for
two profiles of figure \ref{solutions} (hard and soft potential case).
The tail frequency and the central part internal frequency are found relatively close to
the linear frequencies $q_1$ and $q_0$.

\begin{figure}
\psfrag{tau}[0.9]{{\small $t$}}
\psfrag{amp}{{\small$u_1(t)$}}
\begin{center}
\includegraphics[scale=0.3]{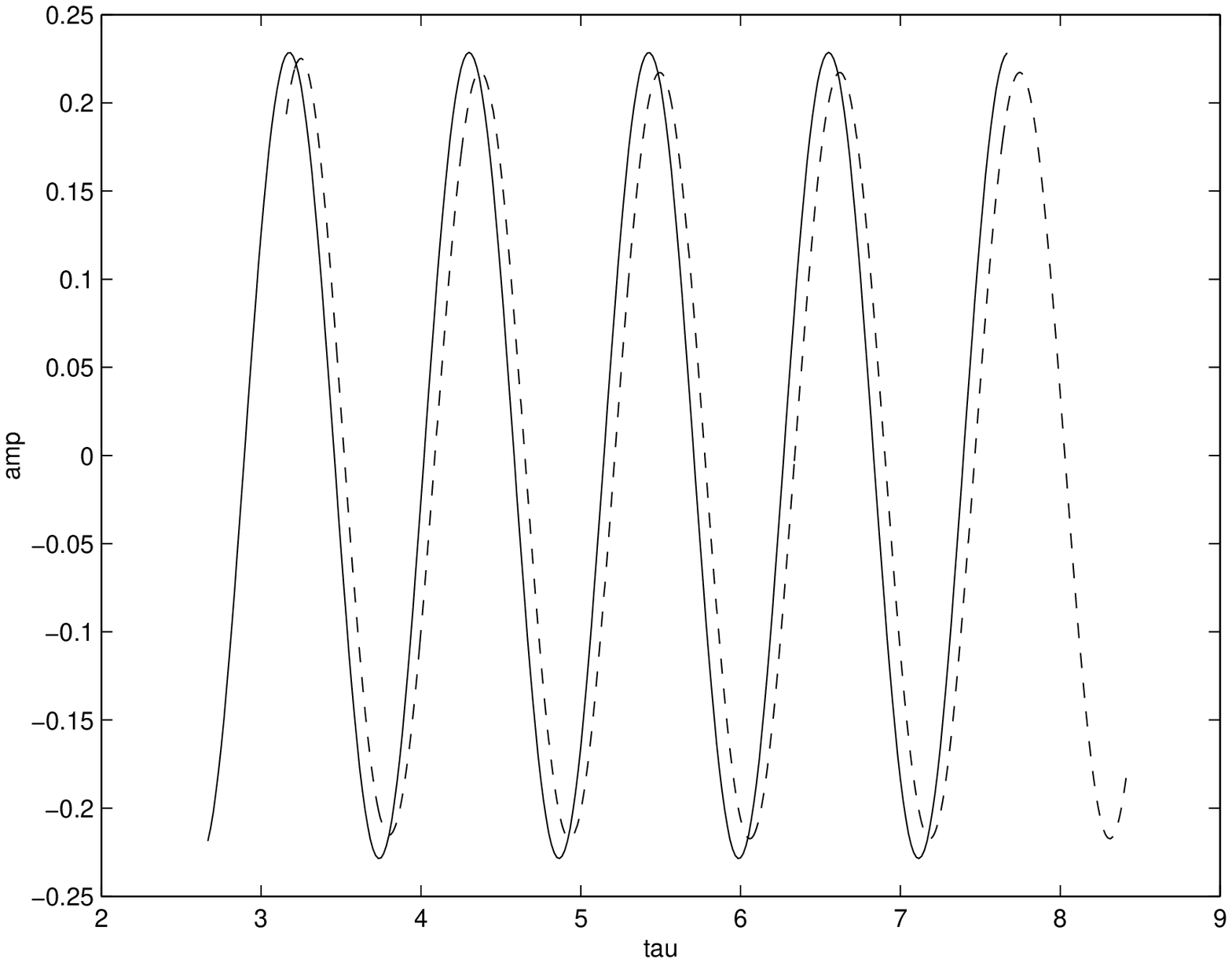}
\end{center}
\caption{\label{comp} Comparison of a nonlinear normal mode issued from the eigenvalues 
$\pm iq_1$ (continuous line) with the tail of
a localized solution of figure \ref{solutions} (dotted line), for $T=5.35$ and 
$V(x)=\frac{1}{2} x^2 +\frac{1}{24} x^4$.
We have $\gamma=\gamma_0 \approx 0.83$, $T_0 \approx 5.59$ and $(T_0,\gamma_0) \in \Gamma_1$.}
\end{figure}

In figure \ref{comp} we compare the tail of a localized solution of figure 
\ref{solutions} with
a numerically computed periodic solution (a nonlinear normal mode) 
originating from the eigenvalues $\pm iq_1$
(the existence of such solutions is guaranteed by the Lyapunov-Devaney theorem).
The tail of the localized solution is very close to the periodic solution, which  
confirms that the large amplitude localized solution 
can be seen as ``homoclinic to periodic'', 
as in the small amplitude regime. 

\vspace{0.1cm}

\begin{figure}
\psfrag{n}[0.9]{{\small $n$}}
\psfrag{xn}{{\small$x_n(0)$}}
\begin{center}
\includegraphics[scale=0.3]{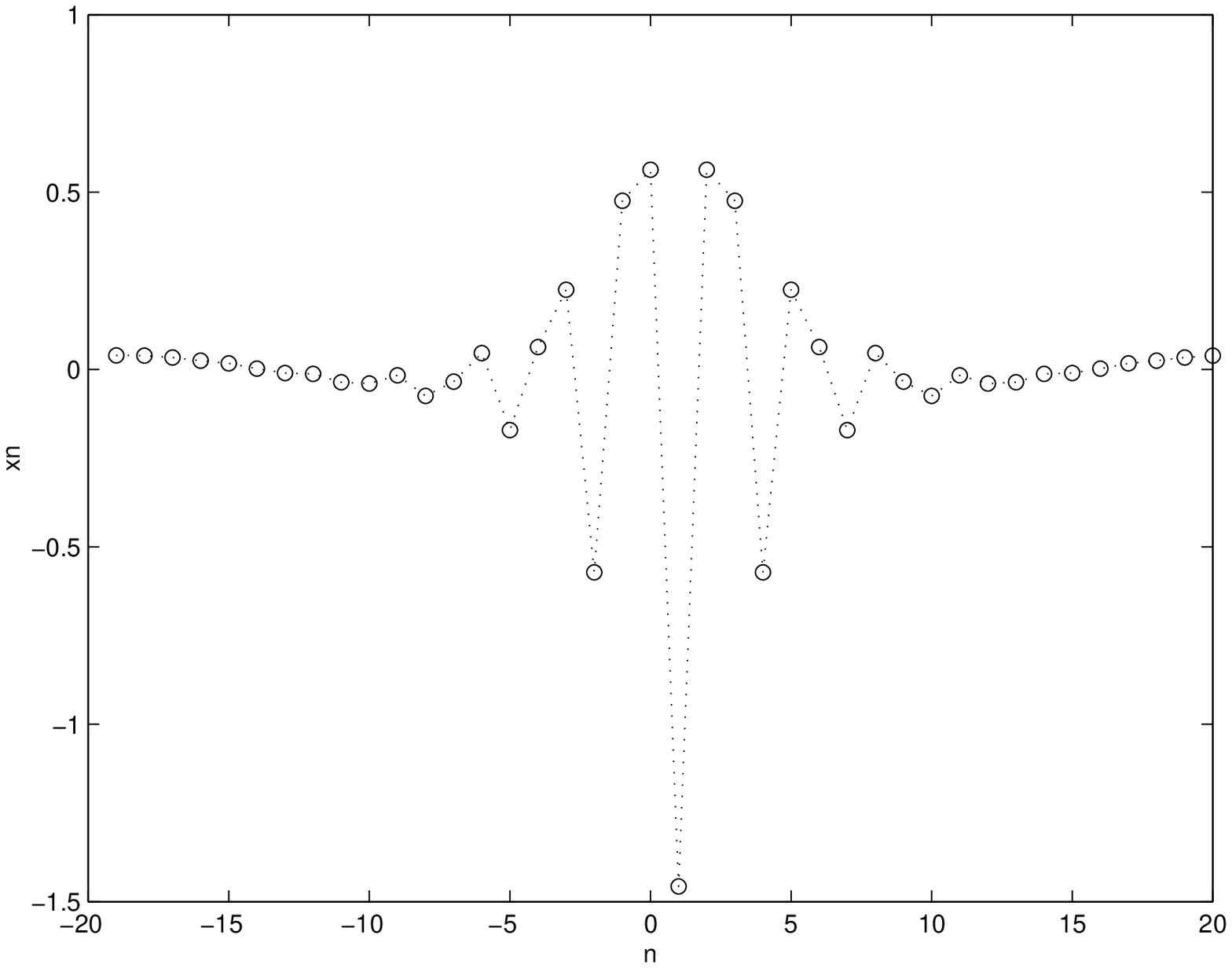}
\end{center}
\caption{\label{reseauHard} Numerical solution of (\ref{eq:KG})-(\ref{def}) for $T=5.5$
and $p=2$.
We consider $V(x)=\frac{1}{2} x^2 +\frac{1}{24} x^4$, 
$\gamma=\gamma_0 \approx 0.83$, $T_0 \approx 5.59$ ($(T_0 ,\gamma_0 ) \in \Gamma_1$).
The solution is plotted as a function of $n$, for $\tau =0$, and corresponds 
via equation (\ref{xnscal}) to a solution
$u_1(t)$ in figure \ref{solutions} (left column, middle).}
\end{figure}

\begin{figure}
\psfrag{n}[0.9]{{\small $n$}}
\psfrag{xn}{{\small$x_n(0)$}}
\begin{center}
\includegraphics[scale=0.3]{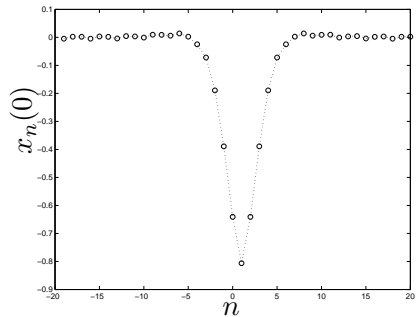}
\end{center}
\caption{\label{reseausoft} Numerical solution of (\ref{eq:KG})-(\ref{def}) for $T=7.4$
and $p=2$. 
We consider $V(x)=\frac{1}{2} x^2 -\frac{1}{4} x^4$, 
$\gamma=\gamma_0 \approx 0.9$, $T_0 \approx 6.63$ ($(T_0 ,\gamma_0 ) \in \Gamma_1$).
The solution is plotted as a function of $n$, for $\tau =0$, and corresponds 
via equation (\ref{xnscal}) to a solution
$u_1(t)$ in figure \ref{solutions} (right column, middle).}
\end{figure}

In figure \ref{reseauHard} we plot as a function of $n$
a travelling breather solution of (\ref{eq:KG}) corresponding 
(via equation (\ref{xnscal})) to
a profile of figure \ref{solutions}. In the case of a hard potential 
one observes that most
neighboring oscillators are out of phase near the breather center.
The case of a soft potential is considered in figure \ref{reseausoft}, 
which shows
that neighboring oscillators are in phase near the center.

\begin{figure}
\psfrag{tau}[0.9]{{\vspace{1cm} \small  \mbox{$\ln \mu$}}}
\psfrag{amp}{{\small $\ln \|u_1\|_{\infty}$}}
\begin{center}
\includegraphics[scale=0.2]{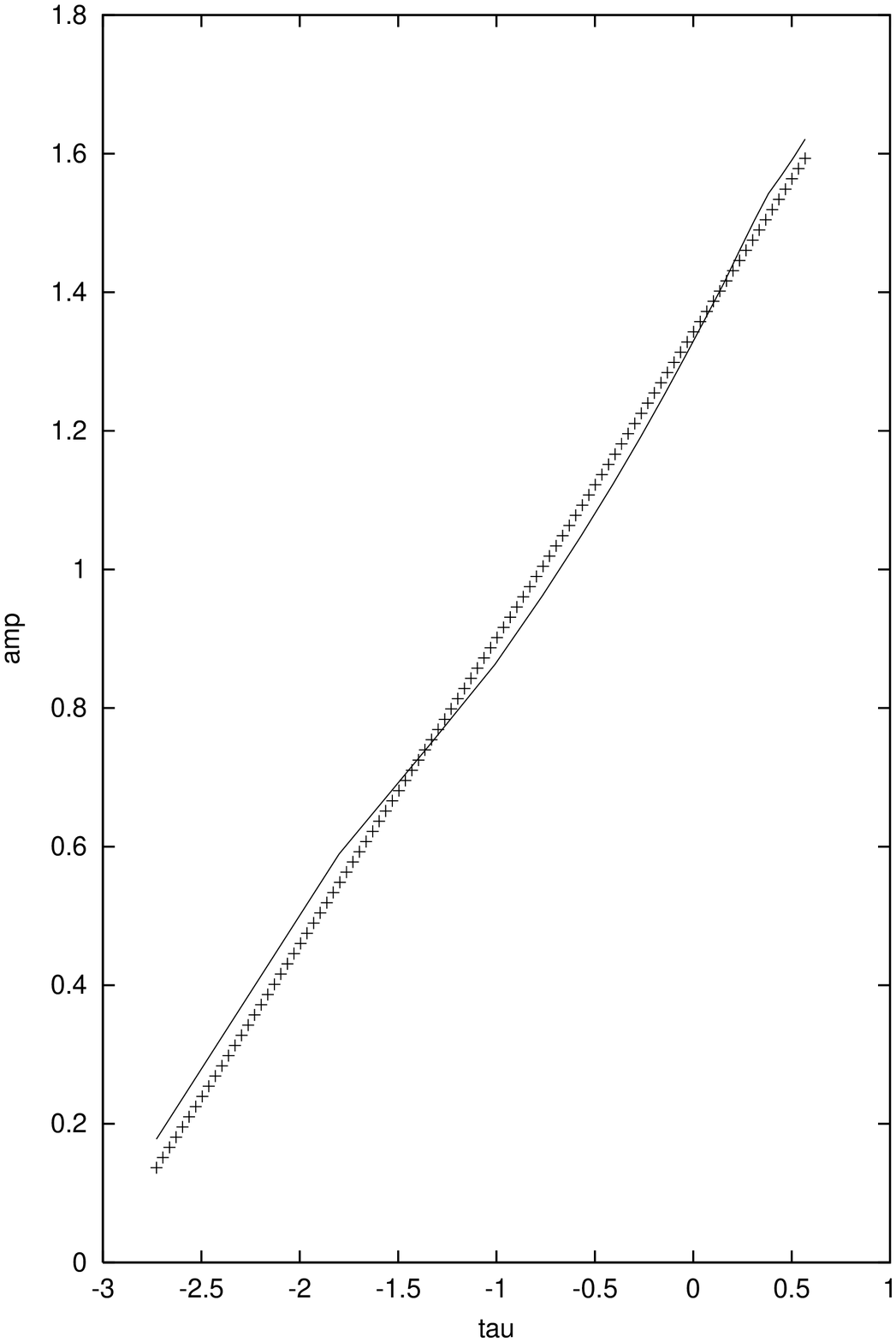}
\includegraphics[scale=0.2]{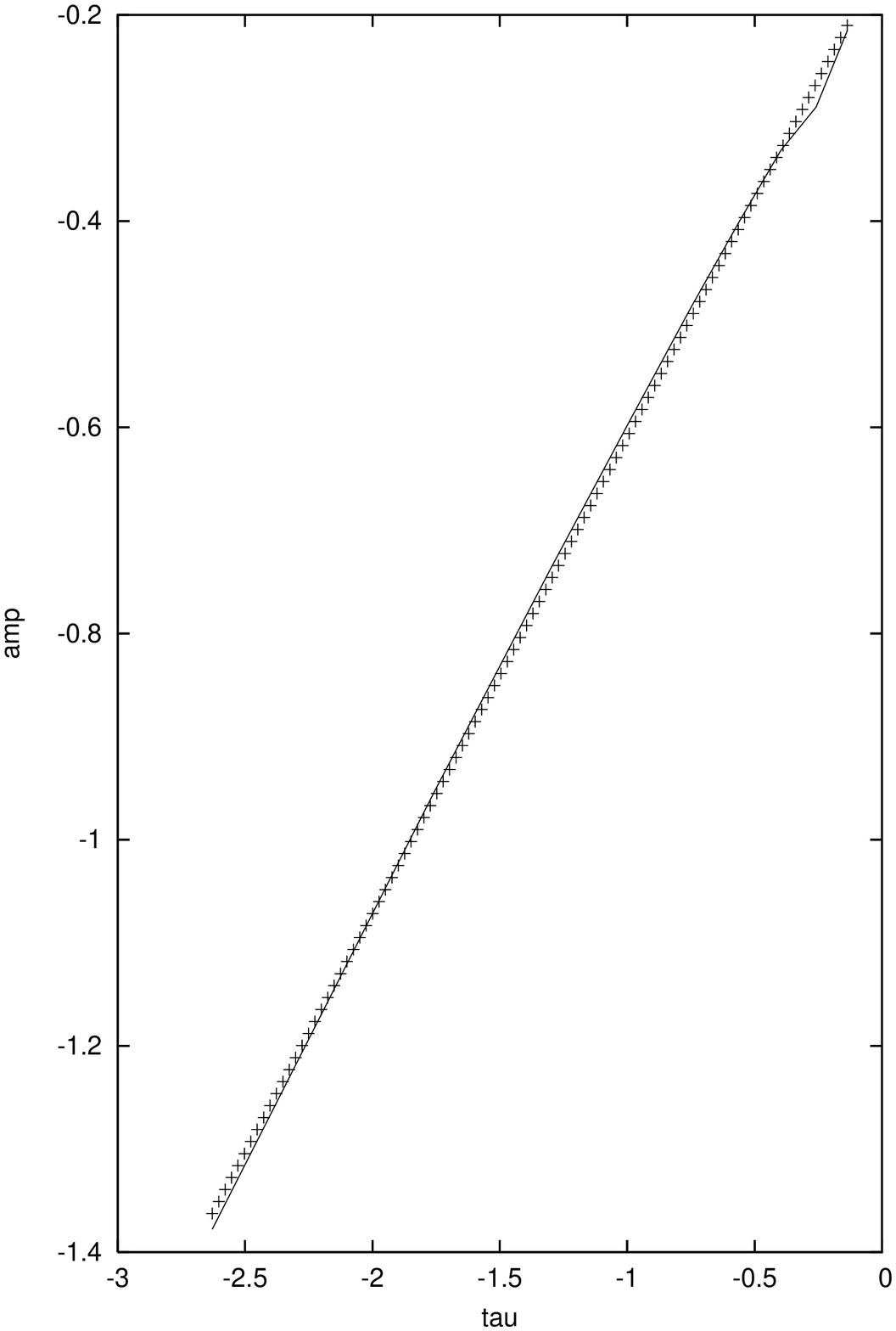}
\end{center}
\caption{\label{ampversusmu} Maximal amplitude $ \|u_1\|_{\infty}$ of a numerical solution of (\ref{scalar}) (continuous line) as a function of $ \mu$ in logarithmic scales. The dotted line represents a linear regression. 
One has $V(x)=\frac{x^2}{2}+\frac{x^4}{24}$, 
$\gamma=\gamma_0 \approx 0.83$, $T_0 \approx 5.59$
for the left figure 
and $V(x)=\frac{x^2}{2}-\frac{x^4}{4}$, $\gamma =\gamma_0 \approx 0.9$, $T_0\approx 6.63$ for the right one. In both cases one has $\mu=|T-T_0|$ and $(T_0,\gamma_0) \in \Gamma_1$.}
\end{figure}

Now let us examine the amplitude of travelling breathers versus parameter $\mu$
in the large amplitude regime. In figure \ref{ampversusmu} we plot 
the amplitude of solutions for polynomial potentials 
$V(x)=\frac{x^2}{2}+\frac{\beta}{4}x^4$. 
One finds that the amplitude behaves approximately as $\mu^\delta $ for 
relatively large values of $\mu$ (corresponding to highly localized solutions),
with $\delta \approx 0.44$ for the hard potential 
(with $\beta =1/6$)
and $\delta \approx 0.46$ for the soft one (with $\beta =-1$). 
In this latter case the amplitude graph becomes less regular as
one enters the next ``tongue'' $\Gamma_2$ (figure \ref{bif}),
which might indicate bifurcations occuring near
the numerically computed solutions
(this part is not visible in figure \ref{ampversusmu}).
 
A $\mu^{1/2}$ dependency of the amplitude (at leading order)
is expected in the small amplitude regime $\mu \approx 0$ \cite{jamessire}.
Surprisingly a rather similar behaviour is observed
in the large amplitude regime.

\begin{figure}
\psfrag{tau}[0.9]{\vspace{0.05cm} {\small $t$}}
\psfrag{amp}{{\small$u_1(t)$}}
\begin{center}
\includegraphics[scale=0.2]{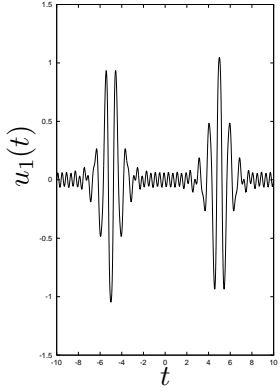}
\end{center}
\caption{\label{double} 
$2-$bumps solution of equation (\ref{scalar}). We consider 
$V(x)=\frac{1}{2}x^2-\frac{1}{4}x^4$, $T_0 \approx 6.63$, $\gamma=\gamma_0 \approx 0.9$ 
($(T_0 ,\gamma_0 ) \in \Gamma_1$) and $T=9$.}
\end{figure}

We have found in addition multibump solutions.
Figure \ref{double} shows e.g. a $2-$bumps solution of equation (\ref{scalar}) 
(note the reverse symmetry between the two bumps). 

\subsection{\label{peror}Modulated periodic solutions}

As previously we consider $(T_0 ,\gamma_0 )\in \Gamma_1 \cap \Delta_0$ 
and choose $\mu=|T-T_0|$ as a small parameter, $\gamma=\gamma_0$ being fixed.
If one neglects higher order terms in the normal form (\ref{normalform}),
the reversible bifurcation near $\Gamma_{2k+1}$ creates
(in addition to homoclinic solutions) a
three-parameter family of periodic solutions
with $C=D=0$ and 
\begin{equation}
\label{perNL}
A_{\mu ,\lambda ,\theta}(t)=r\, e^{i(\Omega t +\theta )}, \ \ \
B_{\mu ,\lambda ,\theta}(t)= i\lambda \, r\, e^{i(\Omega t +\theta )},
\end{equation}
provided one approaches $\Delta$ from below in figure \ref{bif} ($s_1 >0$),
$s_2 <0$ and expression (\ref{perNL}) is considered for
$\lambda \approx 0$ ($\lambda$ is a real parameter).
The frequency $\Omega$ and modulus $r$ depend on $(\mu ,\lambda )$ as follows
\begin{equation}
\label{lambda}
\Omega -q_0 =
\lambda +p_1 -(p_2+2\lambda p_5)\frac{\lambda^2+s_1}{s_2+2\lambda s_5},
\end{equation}
\begin{equation}
\label{defr}
r^2 =-\frac{\lambda^2+s_1}{s_2+2\lambda s_5}.
\end{equation}
Note that coefficients $p_i ,s_i$ depend on $\mu$ and
the first integral $v_4$ of the truncated normal form 
takes the value
$v_4=i(A_{\mu ,\lambda ,\theta}\bar{B}_{\mu ,\lambda ,\theta}
-\bar{A}_{\mu ,\lambda ,\theta}B_{\mu ,\lambda ,\theta})=2r^2\lambda $.
The periodic solutions (\ref{perNL})
are $O(\sqrt{\mu} +|\lambda |)$ as 
$(\mu ,\lambda ) \rightarrow 0$
($s_1$ vanishes on $\Delta$) and their frequency is close to $q_0$
(the colliding pair of double eigenvalues is $\pm iq_0$)
since $p_1$ vanishes on $\Delta$.

These solutions should correspond to solutions of
(\ref{eq:KG}) having the form
\begin{equation}
\label{solappper}
x_n(\tau)\approx  
2(-1)^n\, r\, \cos{[\Omega (\frac{\tau}{T}-\frac{n-1}{2}) +\theta ]}
\end{equation}
at leading order, and
consisting in time-periodic pulsating travelling waves.
One has equivalently
\begin{equation}
\label{soluper}
\left(
\begin{array}[c]{c}
    u_1(t)\\
    u_2(t) 
  \end{array}\right)\approx
2 r\, \cos{[\Omega t +\theta ]}
\left(
  \begin{array}[c]{c}
    -1\\
    1
  \end{array}\right) .
\end{equation}
To obtain exact solutions of (\ref{eq:KG}) instead of 
leading order approximations, one should prove the persistence
of the periodic solutions (\ref{perNL}) as higher order terms
are taken into account in the normal form (\ref{normalform}).
We shall not treat this question in the present paper and leave it
for future works.

For computing the corresponding solutions of 
(\ref{general}) with the same approach as in
section \ref{numcomp}, one has to consider the period $M$
of solutions as a real parameter. 
The ansatz (\ref{soluper}) is used with $\Omega =k\frac{2\pi}{M}$
(for an integer $k\geq 1$)
to initiate the Powell method.
Equations (\ref{lambda}) and (\ref{defr}) determine
$\lambda ,r  \approx 0$ locally 
as functions of the two parameters $\mu\approx 0$
and $M\approx k\frac{2\pi}{q_0}$ (hence $\Omega\approx q_0$). 
If we fix in addition
$M- k\frac{2\pi}{q_0} =O(|\mu |)$, then 
equation (\ref{lambda}) yields $\lambda = O(|\mu |)$
and we obtain $r^2 =-s_1/s_2 +O(\mu^2 )$
using equation (\ref{defr}). Consequently we 
fix $r^2 =-s_1/s_2$ in the ansatz (\ref{soluper}) to initiate the
Powell method near the ``tongue'' $\Gamma_1$.

To adapt the numerical scheme (\ref{num})-(\ref{BC}) to
non-integer values of $M$, it suffices to
rescale equation (\ref{scalar}) into 
\begin{equation}\label{scalarrescale}
\frac{d^2y_1}{dx^2}+M^2T^2V'(y_1)=-M^2\gamma T^2 (y_1(x+\frac{1}{2M})+2y_1(x)+y_1(x-\frac{1}{2M})),
\end{equation}
where $y_1(x)=u_1(t)$, $x=\frac{t}{M}$ and 
(\ref{scalarrescale}) is subject to periodic boundary conditions $y_1(x+1)=y_1(x)$.
One discretizes (\ref{scalarrescale}) with a scheme similar to
(\ref{num})-(\ref{BC}), except 
advance and delay terms are now obtained by linear interpolation.

\begin{figure}
\psfrag{tau}[0.9]{\small $t$}
\psfrag{amp}{\small$u_1(t)$}
\begin{center}
\includegraphics[scale=0.3]{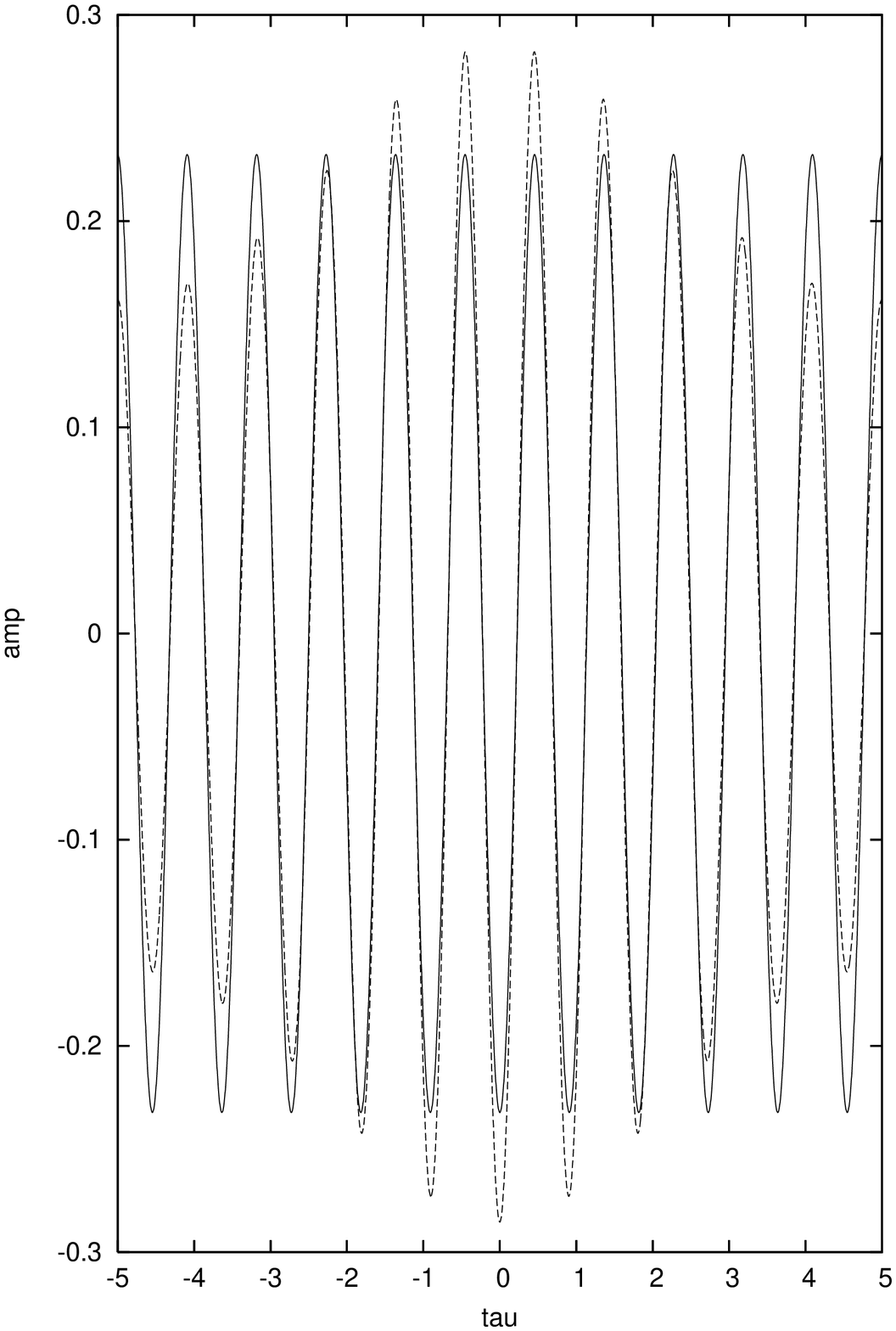}
\includegraphics[scale=0.3]{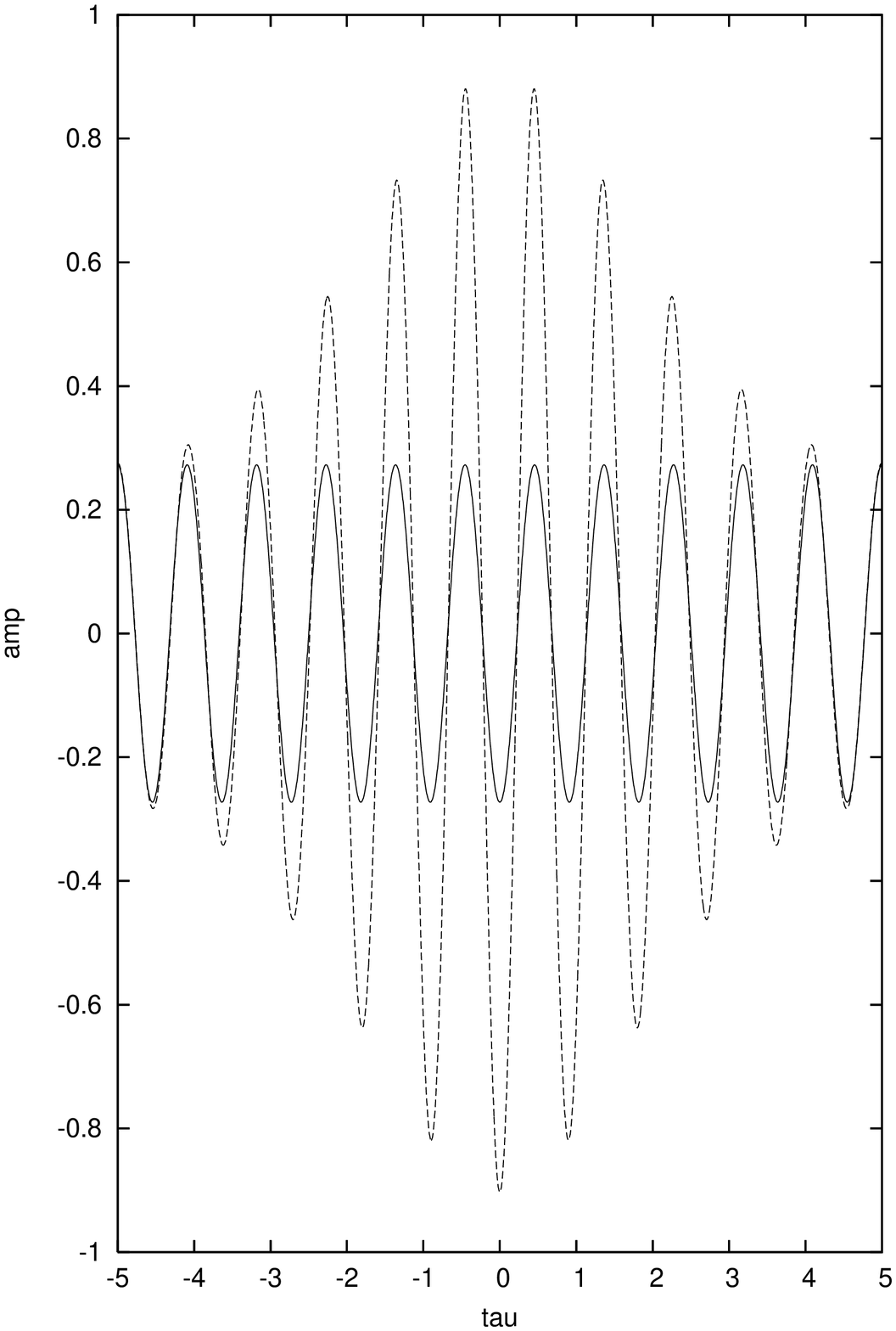}
\includegraphics[scale=0.3]{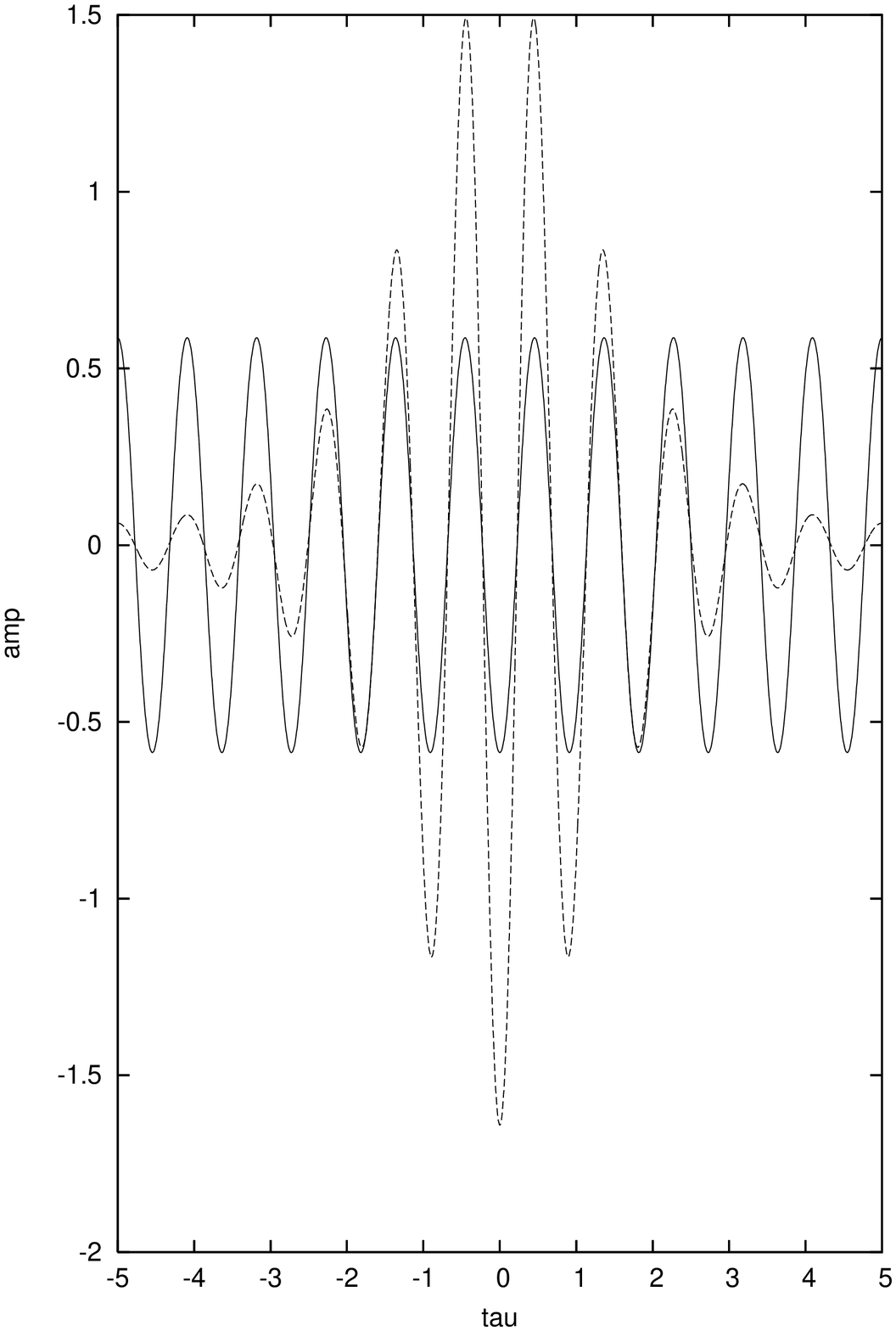}
\end{center} 
\caption{\label{comparaison}Comparison of a periodic solution (continuous line) with a weakly localized one (dotted line) for the  
soft polynomial potential $V(x)=\frac{1}{2}x^2-\frac{1}{4}x^4$
and different values of $T$ : $T=6.75$ (top), $T=6.8$ (middle),
$T=7.2$ (bottom).
We have $\gamma=\gamma_0 \approx 0.9$ and $T_0 \approx 6.63$,
$(T_0 ,\gamma_0 )\in \Gamma_1$.}
\end{figure}

We have numerically observed that localized solutions bifurcate from 
the above family of periodic solutions as 
$M$ is fixed and
$T$ is varied (figure \ref{comparaison}). 
This property could be used to compute localized solutions by continuation from
periodic solutions. 
This method has been previously employed in \cite{feddersen}
for computing travelling breathers in the DNLS system. In this case the problem reduces
to the computation of solitary waves, which bifurcate from families of
(explicitly known) periodic travelling waves. 
This approach seems less efficient in our case since periodic solutions are
not explicitly known and have to be computed numerically.

\vspace{1ex}

In the next section, we present numerical computations in the 
case of a non-even potential, corresponding to system (\ref{eq:system}). 

\section{\label{sec3}Case of non even potentials $V$}

\subsection{\label{approxsol}Approximate solutions}

\begin{figure}
\begin{center}
\includegraphics[scale=0.6]{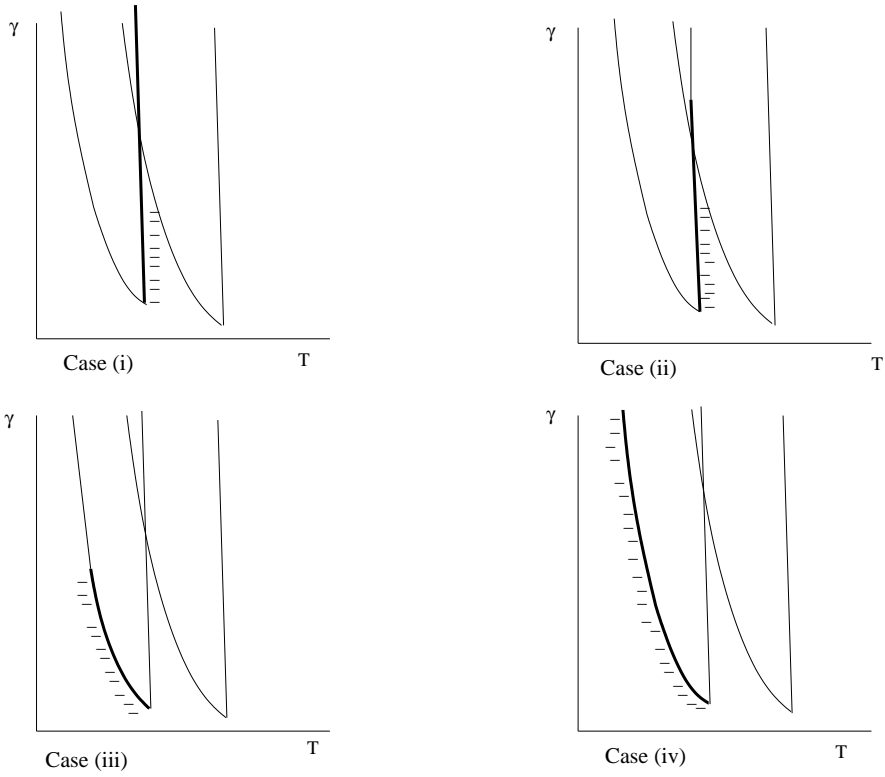}
\end{center}
\caption{\label{homgl} 
Part of $\Gamma_1$ where $s_2 <0$ (bold line).
The dashed area is a small neighborhood of this curve
lying below $\Delta$. 
It corresponds to regions where small amplitude
homoclinic orbits to $0$ bifurcate for the normal form (\ref{normalform})
without higher order terms. The parameter $\kappa$ defined by 
(\ref{defkappa}) varies as follows :  
(i) $\kappa < 10/9$, (ii) $10/9 <\kappa <\kappa_c $ with $\kappa_c \approx 1.15$,
(iii) $\kappa_c < \kappa < 4/3$, (iv) $\kappa > 4/3$.}
\end{figure}

The homoclinic orbits of the truncated normal form
described in section \ref{inguess} 
bifurcate near parts of the ``tongues'' $\Gamma_{m}$
where $s_2 <0$ (see equation (\ref{s2})). 
In the case on non even potentials, these regions
vary with the parameter 
\begin{equation}
\label{defkappa}
\kappa =\beta /\alpha^2,
\end{equation}
depending on the quartic and squared cubic coefficients in the potential $V$
(see (\ref{defv})).
The situation is depicted in figure \ref{homgl} for the curve $\Gamma_1$
(see \cite{jamessire} for the detailed study of the sign of $s_2$).

Now let us assume $s_2(\gamma_0, T_0)<0$ for a fixed 
$(T_0,\gamma_0)\in \Delta_0 \bigcap \Gamma_{2k+1}$ and consider 
$(\gamma,T)\approx (\gamma_0, T_0)$, $(T,\gamma )$ lying below
$\Delta$ in the parameter plane. This is the case e.g. in the dashed
area near $\Gamma_1$ depicted in figure \ref{homgl}.  

Since $V$ is non even, $D=0$ is not in general an invariant subspace
of the full normal form (\ref{normalform}) and 
the oscillatory mode $D(t)$ cannot be eliminated as in section \ref{exact}.
In fact, one expects (this is still a conjecture) that bifurcating
localized solutions have in general nonvanishing 
(up to an exponentially small size)
oscillatory tails $D(t)$, $C(t)$ oscillating at (fast) frequencies 
close to $q_2$, $q_1$. These oscillations are combined with fast oscillations at
a frequency close to $q_0$ (oscillatory part of $A(t)$)
and the slow  hyperbolic dynamics of the envelope of $A(t)$
(time scales are of order $1/\sqrt{\mu}$, $\mu =|T-T_0|+|\gamma -\gamma_0|$).

The normal form (\ref{normalform}) {\it truncated at order $4$} 
admits small amplitude solutions homoclinic to 2-tori (originating
from the pairs of eigenvalues $\pm iq_1 ,\pm iq_2$).
These approximate solutions of (\ref{normalform})
have an explicit form (see \cite{jamessire}, section 5.3).
It has been conjectured in \cite{jamessire} that
they should correspond to the principal part of travelling breather 
solutions of system (\ref{eq:KG}), superposed at infinity on an oscillatory 
(quasiperiodic) tail, and given at leading order by the expression  
\begin{equation}
\label{solappr}
x_n(\tau)\approx
[\,
(-1)^{n}\, A +
(-1)^n\, C +
D
\, ]
\, (\frac{\tau}{T}-\frac{n-1}{2})
+\mbox{c.c.}
\end{equation}
($(A,B,C,D)$ denotes one of the above mentioned approximate solutions of (\ref{normalform}) 
homoclinic to a small quasiperiodic orbit).
The persistence of these homoclinic solutions for the full normal form
is still an open mathematical problem, which would require to generalize
results of Lombardi \cite{lombardi} available when one of the oscillatory modes 
$C(t)$ or $D(t)$ is absent.

\subsection{Numerical computation of travelling breathers}

In this section we consider the Morse potential $V(x)=\frac{1}{2}(1-e^{-x})^2$
and perform the same type of computations as in section \ref{numcomp},
except we now work with system (\ref{eq:system}) 
instead of the scalar equation (\ref{scalar}). 
We consider a point $(T_0 ,\gamma_0 )\approx (6.63,0.9)$ on $\Gamma_1$,
fix $\gamma=\gamma_0$ and let $\mu =T-T_0$ vary.

To initiate the Powell method we use the approximate travelling
breather solutions described in section \ref{inguess}.
For this purpose one has to choose $(T,\gamma )$ near parts of
$\Gamma_{1}$ where $s_2 <0$,
as described in section \ref{approxsol}, figure \ref{homgl}
(dashed area).
The Morse potential has the expansion (\ref{defv}) with
$\alpha =-3/2$, $\beta =7/6$, and $\kappa =14/27$.
  
\begin{figure}
\psfrag{tau}[0.9]{\small $t$}
\psfrag{amp}[1][Bl]{\small $u_1 (t)$}
\begin{center} 
\includegraphics[scale=0.7]{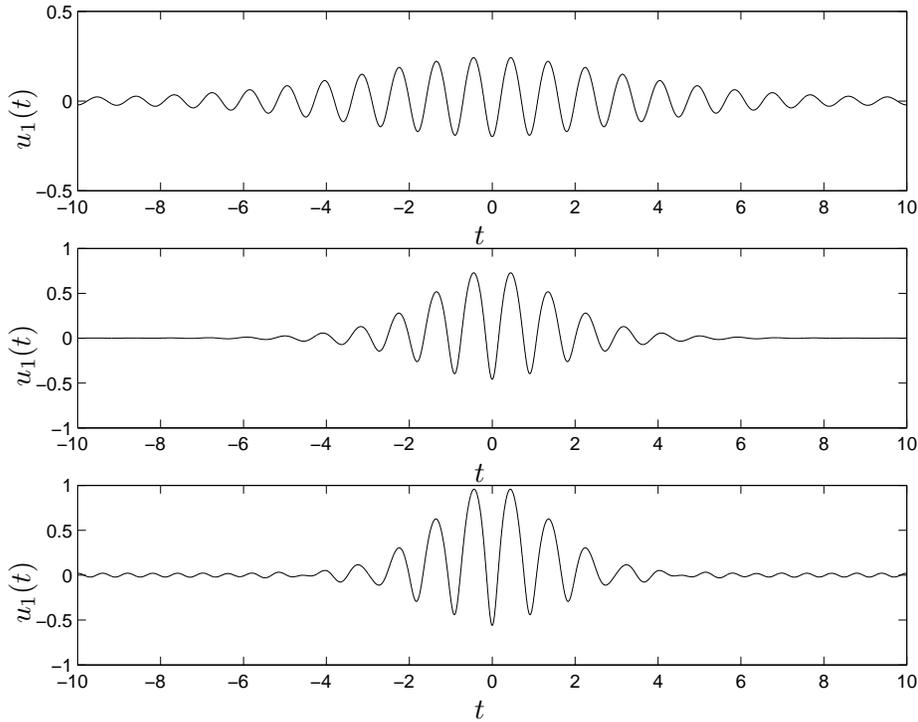}
\end{center} 
\caption{\label{morse} 
Localized numerical solutions of (\ref{eq:system}) 
for various propagation times $T$ and
the Morse potential $V(x)=\frac{1}{2}(1-e^{-x})^2$.
We plot $u_1 (t)$ ($u_2(t)$ behaves similarly).
One has $T=6.7$, $T=7.15$ and $T=7.45$ from top to bottom,
$(T_0 ,\gamma_0 )\approx (6.63 ,0.9)\in \Gamma_1$, $\gamma=\gamma_0$.
The value of $\mu=T-T_0$ increases from top to bottom.}
\end{figure}

Figure \ref{morse} presents numerically computed solutions of (\ref{eq:system})
for different values of $T$. We only plot $u_1 (t)$ because $u_2(t)$ behaves similarly.
As $\mu =T-T_0$ increases, the central hump of the
solution gets more localized and increases in amplitude.

\begin{figure}
\psfrag{tau}[0.9]{\vspace{0.05cm} {\small $t$}}
\begin{center} 
\psfrag{amp}{{\small$u_1(t)$}}
\includegraphics[scale=0.3]{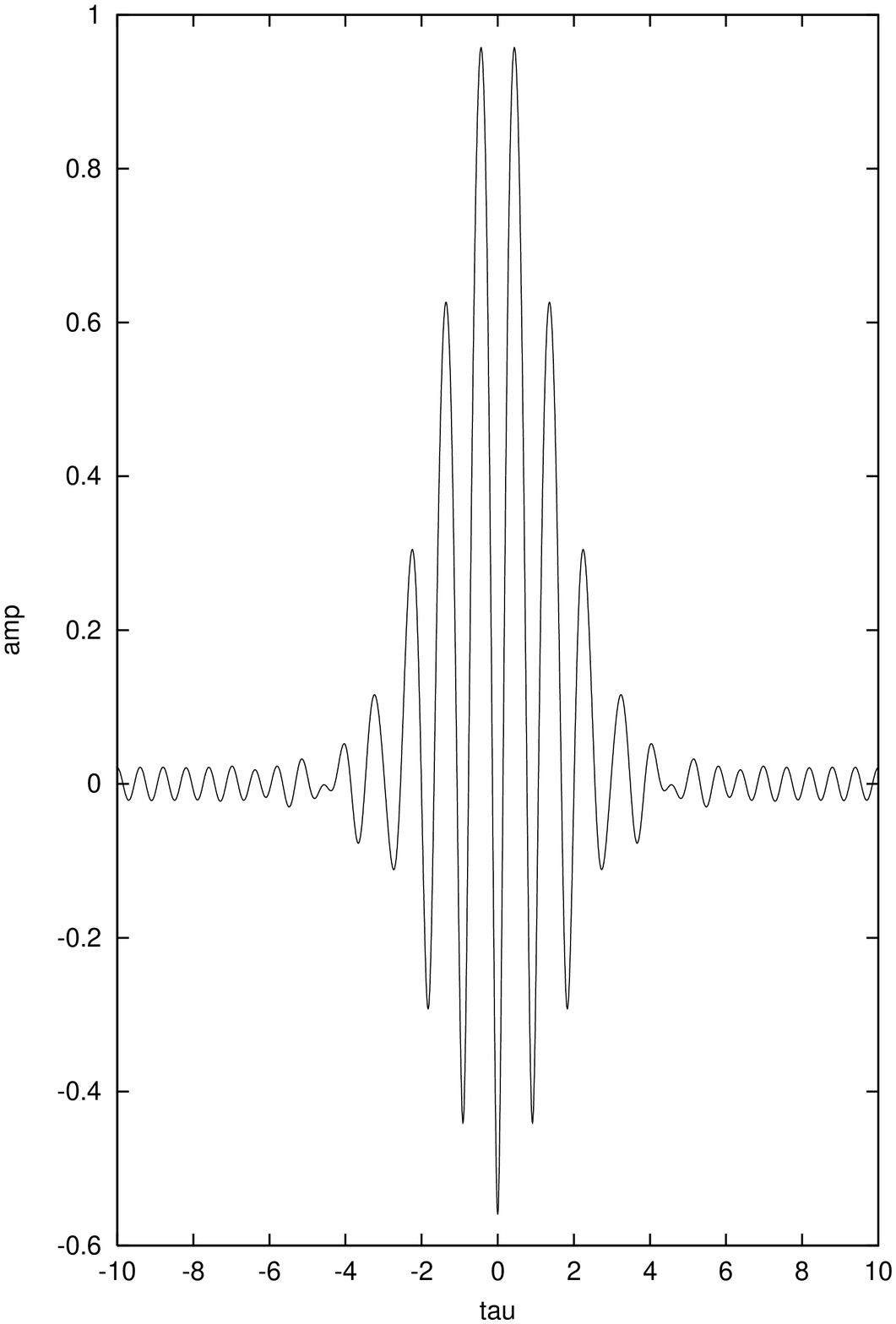}
\psfrag{amp}{{\small$u_2(t)$}}
\includegraphics[scale=0.3]{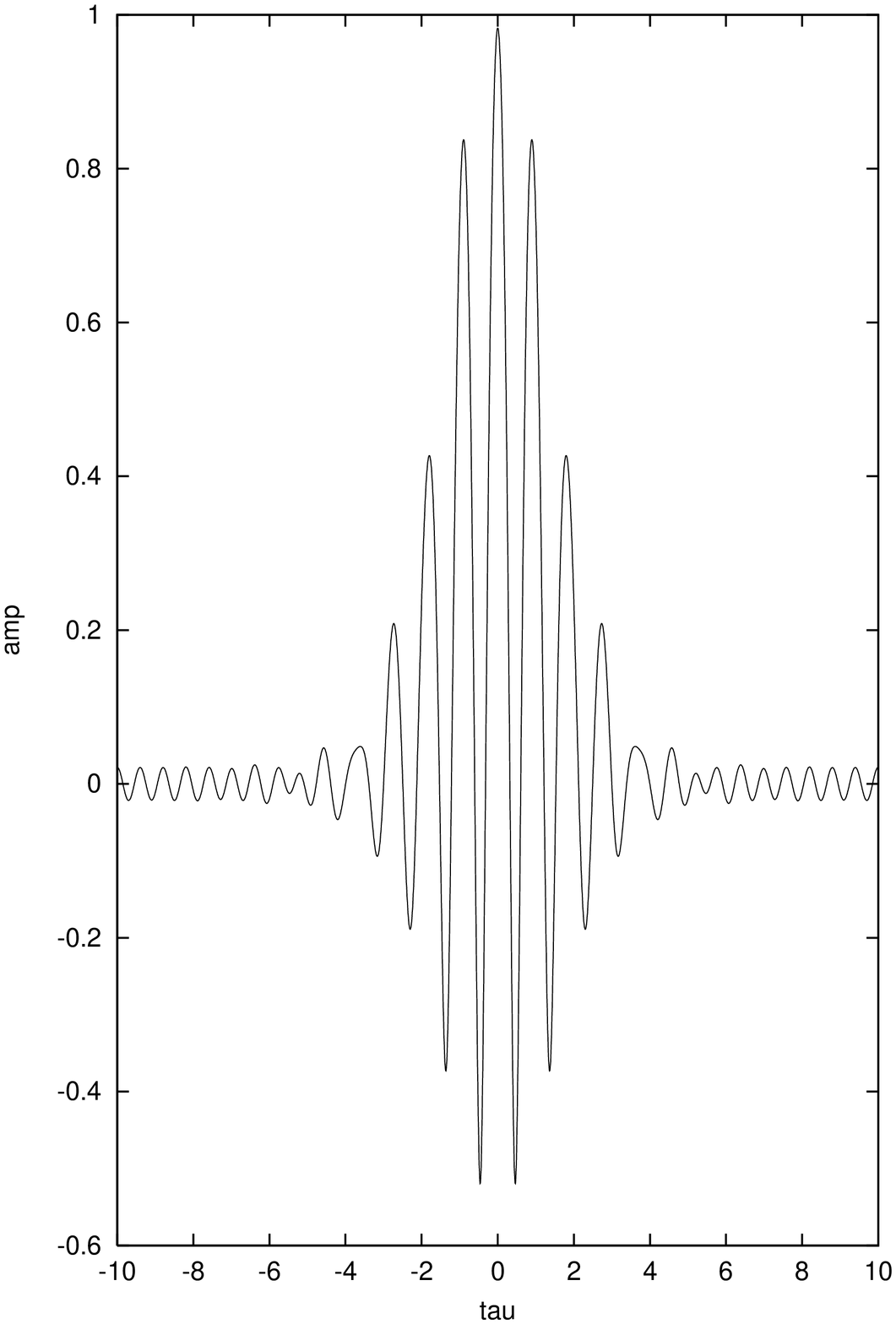}
\end{center} 
\caption{\label{u1u2} 
Localized numerical solution $(u_1 (t),u_2 (t))$ of (\ref{eq:system}) 
for $T=7.45$ and the (asymmetric) Morse potential $V(x)=\frac{1}{2}(1-e^{-x})^2$.
One has $(T_0 ,\gamma_0 )\approx (6.63 ,0.9)\in \Gamma_1$, $\gamma=\gamma_0$.
Note the difference between $u_1 (t)$ and $-u_2 (t)$.}
\end{figure}

Note that $u_1 (t)$ and $-u_2 (t)$ are different, contrarily to the case of
even potentials. This asymmetry can be observed in figure \ref{u1u2}.
It corresponds to the fact that the symmetry
$x_{n+1}(\tau)=-x_n(\tau-\frac{T}{2})$ is broken in the case of non even potentials.

\begin{figure}
\psfrag{n}[0.9]{\small $n$}
\psfrag{xn}[1][Bl]{\small $x_n(0), x_n(T)$}
\begin{center}
\includegraphics[scale=0.3]{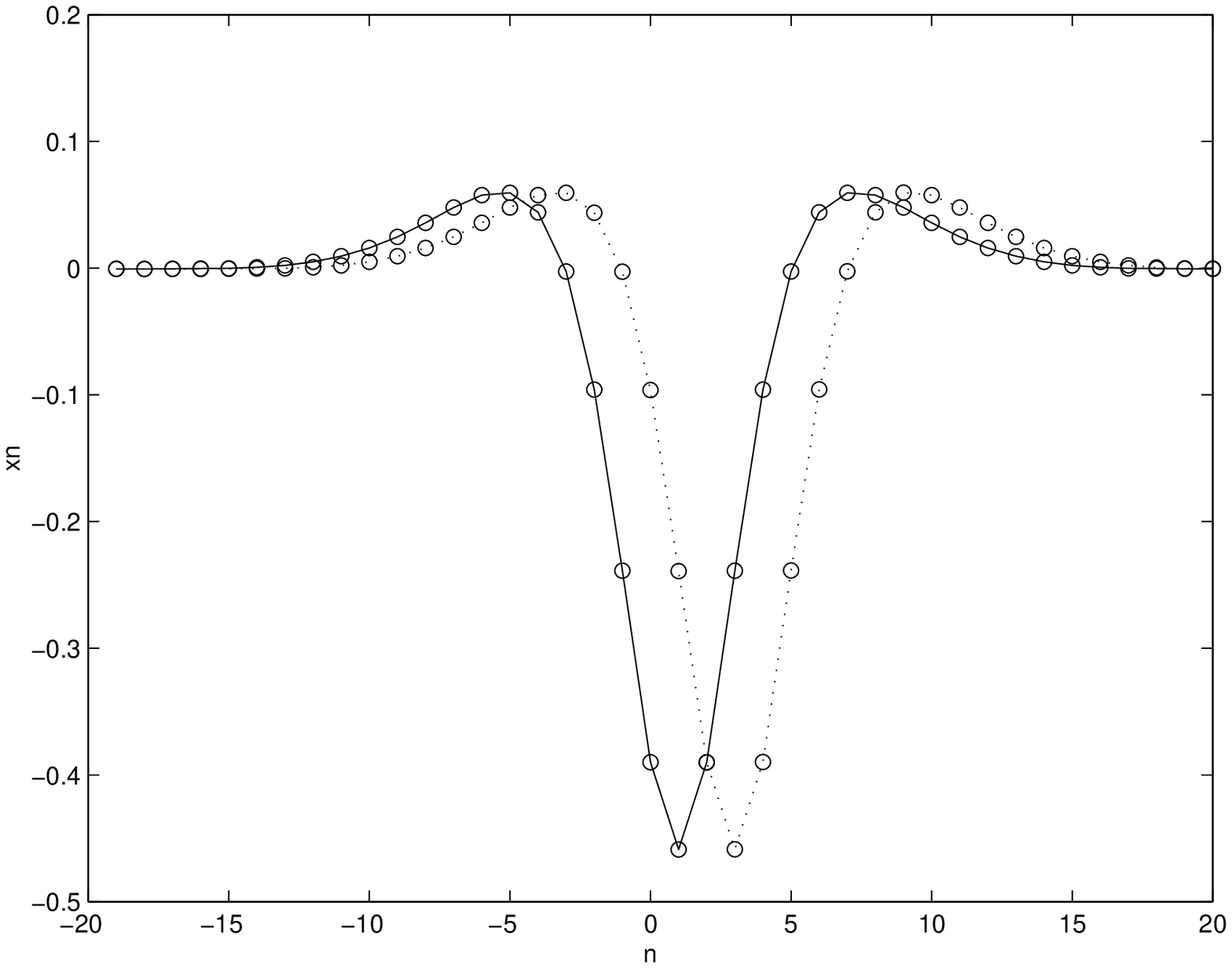}
\includegraphics[scale=0.3]{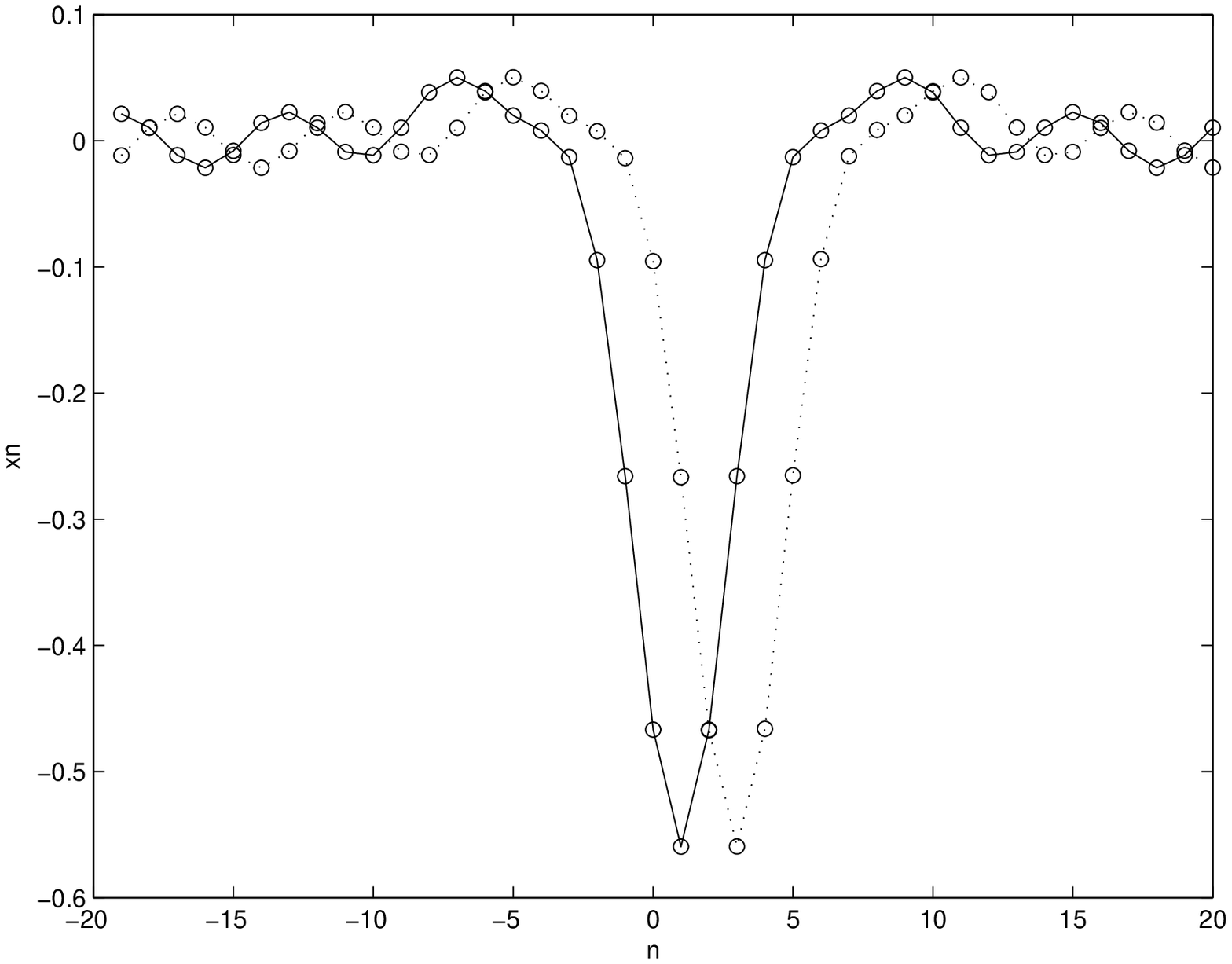}
\end{center} 
\caption{\label{resSansqueue} 
Numerical solution of (\ref{eq:KG})-(\ref{def}) for $T=7.15$ (left), $T=7.45$ (right)
and $p=2$.
We consider $V(x)=\frac{1}{2}(1-e^{-x})^2$,
$\gamma=\gamma_0 \approx 0.9$, $T_0 \approx 6.63$ ($(T_0 ,\gamma_0 ) \in \Gamma_1$).
The solution is plotted as a function of $n$, for $\tau=0$ (continuous line) 
and $\tau =T$ (dotted line). 
It corresponds via system (\ref{eq:system}) to profiles $u_1(t)$ in figure \ref{morse} 
(middle and bottom).}
\end{figure}

In figure \ref{resSansqueue} we plot as a function of $n$
the corresponding mass displacements $x_n (\tau )$, at times $\tau =0$ and $\tau =T$ 
(we fix $T= 7.15$ in the left figure and $T=7.45$ in the right one). 
As prescribed by condition (\ref{def}), 
one recovers the same profile at both times, up to translation by $2$ sites.
One can note that neighboring oscillators are in phase near the
breather center.

At $T=7.15$ (figures \ref{morse}, middle, and figure \ref{resSansqueue}, left)
no tail is visible at least at the figure scale. On the contrary a tail
appears for $T=7.45$ (figures \ref{morse}, bottom, and figure \ref{resSansqueue}, right).
 
\begin{figure}
\psfrag{tau}[0.9]{\small $t$}
\psfrag{amp}[1][Bl]{\small $u_1 (t)$}
\begin{center}
\includegraphics[scale=0.3]{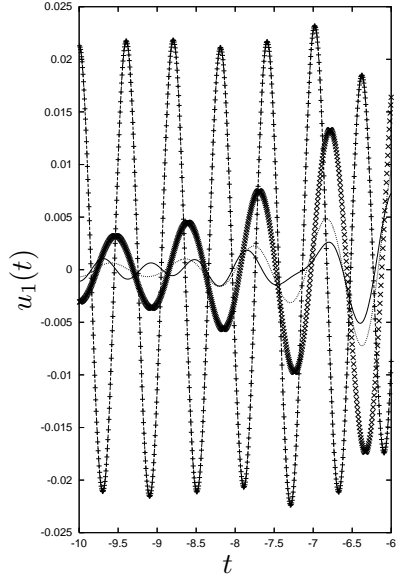}
\end{center} 
\caption{\label{zoommorse} Magnification of the tail of localized
solutions (component $u_1(t)$) 
for the Morse potential with $\gamma=\gamma_0 \approx 0.9$, $T_0 \approx 6.63$
($(T_0, \gamma_0 )\in \Gamma_1 $). 
We consider several values of $T$ : $T=6.9$ (crosses $\times$), 
$T=7.15$ (dashed line),
$T=7.35$ (continuous line), $T=7.45$ (crosses $+$). The minimal tail size is
found around $T=7.15$. 
Cases $T=7.15$ and $T=7.45$ can be compared with figure \ref{morse}.}
\end{figure}

Figure \ref{zoommorse} provides a magnification of the tail of the solution profile.
The oscillations observed near the boundary
for $T=6.9$ are not part of the solution tail and correspond in fact to
oscillations of the central hump at a frequency close to $q_0$
(near this value of $T$ the solution is weakly localized, see figure \ref{morse}, top).
The magnification shows small oscillations for $T=7.15$ with
a frequency close to the frequency observed for $T=6.9$, therefore we attribute
these oscillations to the damped central part of the solution, i.e. to the
component $A(t)$ approximated by (\ref{homsol2}). Consequently we do not find 
for $T=7.15$ a visible track of an
oscillatory tail corresponding to components $C(t),D(t)$.
On the contrary, in figure \ref{zoommorse}
a nondecaying tail with a frequency close to $q_2$ is clearly visible 
at $T=7.45$ and larger values of $T$ (i.e. for larger values of $\mu$).
The nondecaying tail 
is also visible for some smaller values of $T$ but the amplitude becomes very small
(the tail size is decreased approximately by a factor of $100$ for $T=7.35$, 
see figure \ref{zoommorse}).

\begin{figure}
\begin{center} 
\psfrag{omega}[0.9]{\small $ q/ (2\pi ) $}
\psfrag{four}[Bl]{ $|\hat{u}_1|^2$}
\includegraphics[scale=0.3]{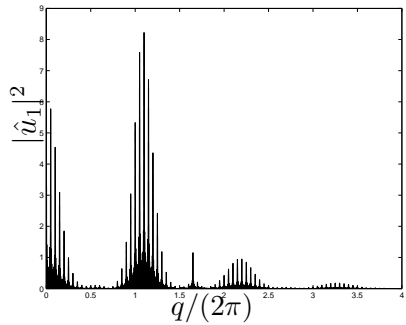}
\end{center} 
\caption{\label{fftmorse} 
Fourier spectrum of $u_1 (t)$ (the spectrum of $u_2(t)$ is similar) for
$T=7.45$, $\gamma=\gamma_0$, $(T_0 ,\gamma_0 )\approx (6.63 ,0.9) \in \Gamma_1$.
The corresponding profile of figure \ref{u1u2} has been extended with periodic
boundary conditions.
Note the sharp frequency peak close to $\frac{q_2}{2\pi} \approx 1.6$.
One has $\frac{q_0}{2\pi} \approx 1.11, \frac{q_1}{2\pi} \approx 2.30$.}
\end{figure}

Figure \ref{fftmorse} shows the Fourier spectrum of $u_1 (t)$ for $T=7.45$
and one can observe a sharp peak at a frequency close to $q_2 /2\pi \approx 1.6$.
The damped oscillations of the central hump
(approximated by (\ref{homsol2})) at a frequency close to $q_0$ appear as
broad peaks and involve harmonics of all orders. 
In particular, figure \ref{morse} (bottom) clearly shows that
the oscillatory part of the central hump has a nonzero time average.
This situation contrasts with
the even potential case (figure \ref{fourier}), in which these
oscillations only involve odd order harmonics.

Another difference with respect to the even potential case is that solutions
tails correspond here to the component $D$ of (\ref{solappr}) instead of $C$
(the tail oscillates at a frequency close to $q_2$ instead of $q_1$).
Consequently the oscillatory tail consists in a time-periodic travelling
wave, instead of a time-periodic pulsating travelling wave as in (\ref{solappr2}).
Note that we conjecture in fact the existence of
a very small oscillatory component $C(t)$ (not visible in figure \ref{u1u2}),
which might be hidden in the broad peak of figure \ref{fftmorse} close
to $2q_0$ ($\frac{2q_0}{2\pi} \approx 2.22$ and $\frac{q_1}{2\pi} \approx 2.30$
are rather close). This is by analogy to the small amplitude regime described
by the normal form (\ref{normalform}). Indeed we conjecture that bifurcating
homoclinic orbits of the full normal form have generically
nonvanishing components $C,D$, except if $T,\gamma $ satisfy two
(one for each component) compatibility conditions
(\cite{jamessire}, section 5.3).  

\begin{figure}
\psfrag{tau}[0.9]{{\small $\ln \mu$}}
\psfrag{amp}{{\small$\ln \|u_1\|_{\infty}$}}
\begin{center}
\includegraphics[scale=0.3]{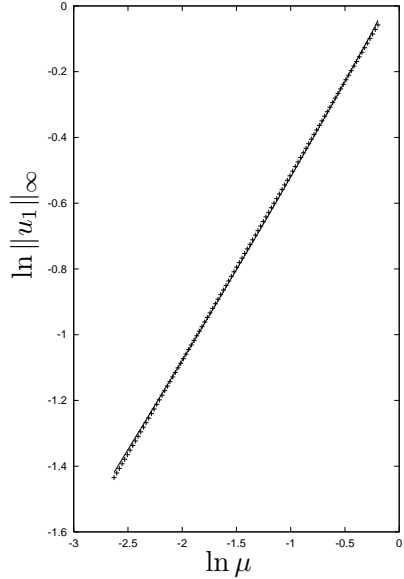}
\end{center}
\caption{\label{vcmorse} 
Maximal amplitude $\|u_1\|_{\infty}$ of a numerical solution of (\ref{eq:system}) (continuous line) as a function of $\mu$, in logarithmic scales. The dotted line represents a linear regression. One has $V(x)=\frac{1}{2}(1-e^{-x})^2$, $\gamma=\gamma_0 \approx 0.9$, $T_0\approx 6.63$ 
($(T_0,\gamma_0) \in \Gamma_1$), 
$\mu=T-T_0$. A similar behaviour is observed for $u_2(t)$.}
\end{figure}

We end by a study of travelling breathers amplitude versus $\mu=T-T_0$
in the large amplitude regime (figure \ref{vcmorse}). 
One finds that the amplitude behaves approximately as $\mu^\delta $ for 
relatively large values of $\mu$ (corresponding to highly localized solutions),
with $\delta \approx 0.56$. 
A $\mu^{1/2}$ dependency of the amplitude (at leading order)
is expected in the small amplitude regime $\mu \approx 0$ \cite{jamessire}.
Surprisingly a rather similar behaviour is observed
in the large amplitude regime. In addition the scaling in amplitude
is rather similar to the case of equation (\ref{scalar}) with a
quartic interaction potential (section \ref{numcomp}).

\end{document}